%% file: astraea.tex
\documentclass[10pt,conference]{IEEEtran}
\IEEEoverridecommandlockouts
\usepackage{amsmath,amssymb,amsfonts}
\usepackage{tikz}
\usepackage[normalem]{ulem}
\usepackage{caption}
\usepackage{algorithm} 
\usepackage{xcolor}
\usepackage{cleveref}
\usepackage{multirow}
\usepackage{tabularx}
\usepackage{subfig}
\usepackage{cite}
\usepackage{makecell}
\usepackage{booktabs}
\usepackage{cancel}
\usepackage{xurl}
\usepackage[subtle]{savetrees}
\usepackage{enumitem}

\crefformat{section}{\S#2#1#3} 
\crefformat{subsection}{\S#2#1#3}
\crefformat{subsubsection}{\S#2#1#3}
\definecolor{mertcolor}{RGB}{145,26,27}

\newcommand*\circled[1]{\tikz[baseline=(char.base)]{
            \node[shape=circle,draw=mertcolor,inner sep=1pt, text=mertcolor] (char) {#1};}}

\def\BibTeX{{\rm B\kern-.05em{\sc i\kern-.025em b}\kern-.08em
    T\kern-.1667em\lower.7ex\hbox{E}\kern-.125emX}}
\begin{document}

\newcommand{\tool}{Astraea}

\title{An Online Probabilistic Distributed Tracing System}



\author{
\IEEEauthorblockN{M. Toslali\IEEEauthorrefmark{1}, S. Qasim\IEEEauthorrefmark{1}, S. Parthasarathy\IEEEauthorrefmark{2}, F. A. Oliveira\IEEEauthorrefmark{2}, H. Huang\IEEEauthorrefmark{2}, G. Stringhini\IEEEauthorrefmark{1}, Z. Liu\IEEEauthorrefmark{3}, A. K. Coskun\IEEEauthorrefmark{1}}
\IEEEauthorrefmark{1}Boston University, \IEEEauthorrefmark{2}IBM Research,\IEEEauthorrefmark{3}University of Maryland
}



\maketitle

\begin{abstract}
  Distributed tracing has become a fundamental tool for diagnosing performance issues in the cloud by recording causally ordered, end-to-end workflows of request executions. However, tracing in production workloads can introduce significant overheads due to the extensive instrumentation needed for identifying performance variations.
This paper addresses the trade-off between the cost of tracing and the utility of the ``spans" within that trace through \tool, an online probabilistic distributed tracing system. \tool~is based on our technique that combines online Bayesian learning and multi-armed bandit frameworks. This formulation enables \tool~to effectively steer tracing towards the useful instrumentation needed for accurate performance diagnosis.
\tool~ localizes performance variations using only 10-28\% of available instrumentation, markedly reducing tracing overhead, storage, compute costs, and trace analysis time.
  
  
\end{abstract}
\begin{IEEEkeywords}
Distributed Systems, Performance Diagnosis, Microservices, Cloud Computing, online Bayesian learning.
\end{IEEEkeywords}

\section{Introduction}


  \label{sec:intro}
\input{intro.tex}

\section{Background and related work}
\label{sec:rel}
\input{rel.tex}

\section{\tool}
\label{sec:design}
\input{design.tex}
\section{Experimental evaluation}
\label{sec:experiments}
\input{eval.tex}


\section{Conclusion}

We presented \tool, an online, probabilistic tracing system designed to combat the cost vs. utility of the ``spans" trade-off. Our key contribution was formulating this tradeoff as an exploration vs. exploitation problem. Unlike prior work, \tool~embodies a statistically rigorous, online approach to accurately and adaptively localize and trace the sources of varying performance variations. Further, we demonstrated \tool's practicality, accuracy, and resource efficiency through extensive experiments. We showed that \tool~ can localize 92 \% of the performance variations using only 25\% of the available instrumentation in three popular distributed cloud applications (Social network, Media, and Train ticket), while existing approaches, Log2 and VAIF, achieve 65\% and 71\% accuracy, incurring 3-24x more latency overhead. As more complex services will continue to be built in the cloud, systems like \tool~ can help localize performance problems without introducing significant overheads.

\bibliographystyle{ieeetr}
\bibliography{sample-base}










\end{document}

%% file: intro.tex
Performance variations constitute a common challenge in the cloud, and diagnosing them can be a time-consuming process~\cite{akamai:perf,firm,tailatscale,canopy,sambasivan2011diagnosing,jackpot,iter8,ardelean2018performance,xtrace,alibaba,perfdebug,realworld,xray,analysis,transient,taming}.
For instance, diagnosing an unexpected slowdown in a request can consume hours or even days~\cite{trainticket}.
Distributed tracing has emerged as an essential tool for diagnosing performance variations in the cloud~\cite{sambasivan2011diagnosing,sambasivan2016principled,vaif, delimitsage,delimitseer,pivot,canopy,firm,xtrace,sifter,dapper, jaeger}.
Distributed tracing enables tracking the journey of a request within a distributed cloud application, as it moves through various services.
A \textit{trace} consists of instrumentation choices known as \textit{spans}, which capture causal relationships and the propagation of latency from downstream to upstream operations.
The end-to-end narrative of a request provided by distributed tracing reveals \textit{what went wrong}, making it 
easier to pinpoint and resolve the issues~\cite{delimitsage,delimitseer, jackpot,sambasivan2011diagnosing,sambasivan2016principled,vaif,pivot,canopy,firm,ardelean2018performance,xtrace,sifter,universal-mace,dapper,zhang2009precise}. 

\begin{figure*}[t!bhp]%
    \centering
    \subfloat[Tracing overhead on mean ($\Delta\mu$) and tail ($\Delta.99$) latency \label{fig:tracingoverhead}]{{\includegraphics[width=0.74\columnwidth]{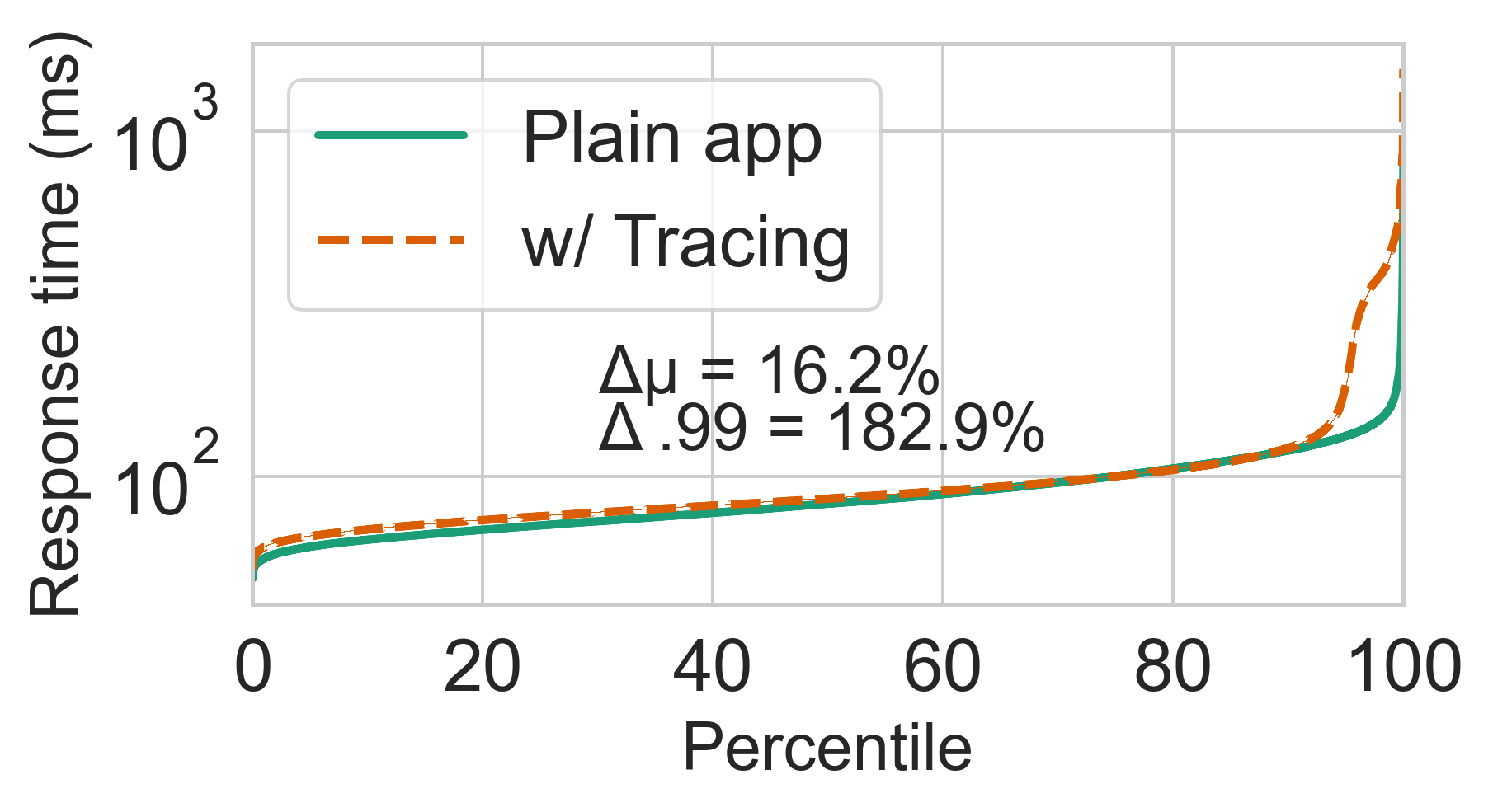} }}
    \subfloat[Cumulative latency variance of spans\label{fig:tracingvariance}]{{\includegraphics[width=0.77\columnwidth]{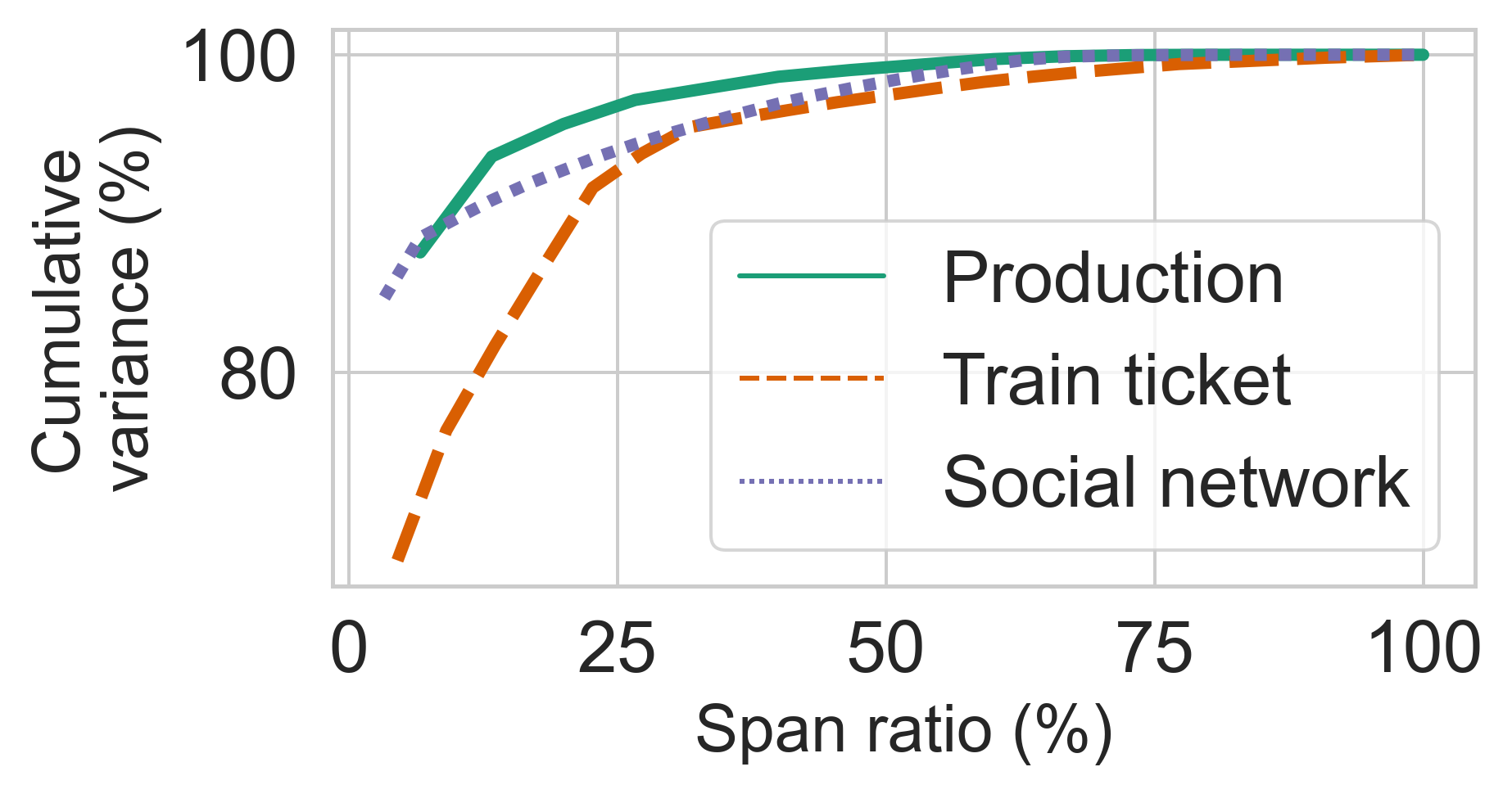} }}
    \caption{The practicality of tracing is limited by (a) overhead and (b) large portion of extraneous instrumentation. (a) Tracing overheads can increase end-to-end request latency. Tracing in this experiment is conducted with a 100\% sampling rate to emphasize the overhead. (b) The majority of spans in a trace are extraneous to explain variation.}%
    \label{fig:tracingmotivation}%
\end{figure*}

\vspace{0.05in}
\noindent\textbf{Tracing cost vs. utility trade-off.}
Predicting which code locations to instrument to diagnose potential future problems is a challenging task~\cite{vaif,log2}.
Developers aim to trace all potential application behaviors to enable diagnosis of new problems.
However, 
recording, processing, and storing traces from production workloads result in significant overheads in terms of storage, computation, and network usage~\cite{canopy,hindsight}.
For instance, Facebook reported generating 1.16 GB/s of tracing data~\cite{canopy}, necessitating substantial backend infrastructure and storage capacities~\cite{hindsight}.
Furthermore, the computational overhead of tracing can lead to an increase in end-to-end request latency, which is not tolerable for end users~\cite{tracinginpractice}. 
Fig. \ref{fig:tracingoverhead} shows the response time overhead of tracing (i.e., Jaeger~\cite{jaeger}) imposed on the Train ticket microservice application~\cite{trainticket}, revealing 180\% and 16\% overhead on the tail and average latency, respectively, 
observed in our experiments. Similarly, Google reported an average response time increase of 16.3\% in their search engine due to tracing~\cite{dapper}.

Distributed tracing often employs request-based sampling as a means to mitigate costs, allowing for tracing only a subset of requests~\cite{jaeger,sifter,tracinginpractice}.
While request-based sampling partially addresses the overhead problem, when it comes to performance diagnosis, the key insight is that majority of spans in a sampled request are extraneous to explain performance variation under investigation~\cite{vaif,log2,pivot,canopy,trainticket}.
For instance, an analysis of the Alibaba traces reveals that 90\% of the spans provide no useful information, with only 10\% of spans being essential for performance diagnosis~\cite{alibaba}.
Fig. \ref{fig:tracingvariance} substantiates these observations by illustrating the cumulative variance of spans from Train ticket~\cite{trainticket}, Social network~\cite{delimit-deathstar}, and a large-scale Internet application (denoted as Production in Fig. \ref{fig:tracingmotivation}), showing that 15\% of spans account for over 80\% of the total variance.
Identifying this ``vital" set of instrumentation with the most utility is a challenge as this information is unknown \textit{a priori} or even changes over time. 
For example, researchers revealed that 28K revisions in Hadoop, HBase, and ZooKeeper applications were made only to insert or modify instrumentation choices concerning cost vs. utility trade-off~\cite{log20}.

\vspace{0.05in}
\noindent\textbf{Automated control of instrumentation.}
To address the tracing cost vs. utility trade-off, ideally, a tracing system should adaptively and automatically control instrumentation choices to enable vital ones needed to diagnose performance variations~\cite{vaif,log2}.
The key requirements of such a system are:
(R1) correctly identifying the vital instrumentation needed to explain on-going performance variations,
(R2) adapting to new sources of performance variations that might evolve over time,
(R3) providing a low-overhead and scalable mechanism to efficiently control instrumentation, 
and (R4) addressing all requirements in a practical and developer-friendly manner. 

Researchers have designed automated techniques to selectively enable the instrumentation needed (\cref{sec:rel}).
VAIF~\cite{vaif} and Log\textsuperscript{2}~\cite{log2} are two automated instrumentation systems that are tailored for performance diagnosis. 
Log\textsuperscript{2} captures all logging records in individual processes and persists ones that contribute to performance variation.
However, it inherently lacks the causal context provided by distributed tracing, which captures latency propagation among distributed services.(\cancel{R1} and \cancel{R2}).
Its tail-based approach, focused on persisting logging records post-execution, primarily reduces storage overheads but neglects the computational overheads (\cancel{R3}).

VAIF is an automated tracing system tailored for performance diagnosis of distributed cloud applications. In an offline profiling phase, VAIF memorizes execution paths of requests through exhaustive workloads~\cite{vaif}. When triggered by developers at runtime, it enables spans within the region of code that currently exhibits the highest variance. VAIF is effective in localizing sources of performance variations, however, considerable developer effort is needed to run comprehensive workloads \textit{a priori} to collect representative traces for VAIF, which is further exacerbated by the need for recurring developer-driven profiling efforts for each new code delivery (\cancel{R4}).
Second, VAIF's span decisions are binary, offering only options of enabling or disabling spans. While this simplifies decision-making and implementation, completely disabling spans may result in a loss of accuracy as no new observations are captured for ongoing or new performance variations, as shown in~\cref{sec:comparison} (\cancel{R1} and \cancel{R2}).

\vspace{0.05in}
\noindent\textbf{Our work.}
We present \tool, a system backed by statistical rigor to accurately and adaptively control instrumentation to address the cost vs. utility trade-off of distributed tracing in performance diagnosis.
\tool~works on top of request-based sampling to further reduce the size of individual traces utilizing span-level sampling probability.
\tool~operates online from the start, eliminating the need for time-consuming offline phases. It begins with the same default level of instrumentation as current distributed tracing practices. As \tool~learns, it gradually decreases the sampling probability of unnecessary instrumentation, leading to continuous performance improvement.
\tool~employs a probabilistic approach, enabling nuanced decision-making by sampling spans with varying probabilities for efficient resource utilization, accurate coverage, and adaptability to changing execution patterns or new issues.

\tool~formulates the cost vs. utility trade-off as an exploration vs. exploitation problem~\cite{Sutton1998}. 
It continuously learns spans that explain performance variations (explore) and dynamically steers tracing toward them (exploit).
This formulation enables an online learning setting tailored for automated control of instrumentation to enhance tracing utility.
The Bayesian framework of \tool~enables space- and time-efficient computation by mapping large trace data to a low-dimensional Bayesian belief representation, gradually building beliefs of spans, thus eliminating the need for extensive trace data storage (cf. \cref{sec:design}).
Based on accumulated beliefs, \tool~employs a probability matching decision strategy where the sampling probability of a span is proportional to the fraction of observations that a span emerges as a vital contributor to variation, aiding in continuously exploring and resolving uncertainty to identify vital spans~\cite{thompson:sampling}.

Overall, \tool~allows
more accurately and adaptively enabling spans needed to explain ongoing as well as new performance variations (R1 and R2), 
incurring low overheads and scaling well to large production traces (R3), and 
circumventing cumbersome offline phases of training or retraining (R4).
\tool~ aims to unleash the power of distributed tracing to an unprecedented practical level by enabling developers to freely instrument their
source code during development, exempt from the headache of cost and manual analysis of a large number of extraneous spans in response to performance variations at runtime. 

\vspace{0.05in}
\noindent\textbf{Our contributions and highlights.}

\begin{itemize}[leftmargin=*]
  \item We devise and formulate the tracing cost vs. utility trade-off as an exploration vs. exploitation problem, enabling an online learning setting. Building on our formulation, we design and implement \tool, 
  which dynamically steers tracing toward vital regions. 
  \item We show that \tool~ accurately localizes injected performance variations 92\% of the time, using only 25\% of instrumentation in three widely used distributed cloud applications (Social network, Media, and Train ticket) (\cref{sec:experiments}). In contrast, Log\textsuperscript{2} and VAIF achieve 65\% and 71\% accuracy, enabling comparable or larger instrumentation and incurring 3-24x more latency overhead. 
  \item We show that \tool's learning framework is fast and scalable. Using traces from a large-scale Internet company, we highlight that \tool's inference duration stays below $100ms$ even for the largest traces, which involve a higher number of spans (\cref{sec:production}).
  \item We demonstrate the efficacy of \tool~ in diagnosing 
  performance variations in Social network, Media, and Train ticket applications. 
  \tool~ pinpoints sources of high variance due to implementation bugs, resource-related issues, network delays, deployments, and inefficient instrumentation, enabling only 11-28\% of available instrumentation (\cref{sec:cases}). 
\end{itemize}





%% file: rel.tex
This section provides a background on distributed tracing and reviews prior work on tracing, instrumentation, and their automation.

\begin{figure}[b]
\includegraphics[trim={0.28cm 0.1cm 0.88cm 1.18cm},clip,width=\columnwidth]{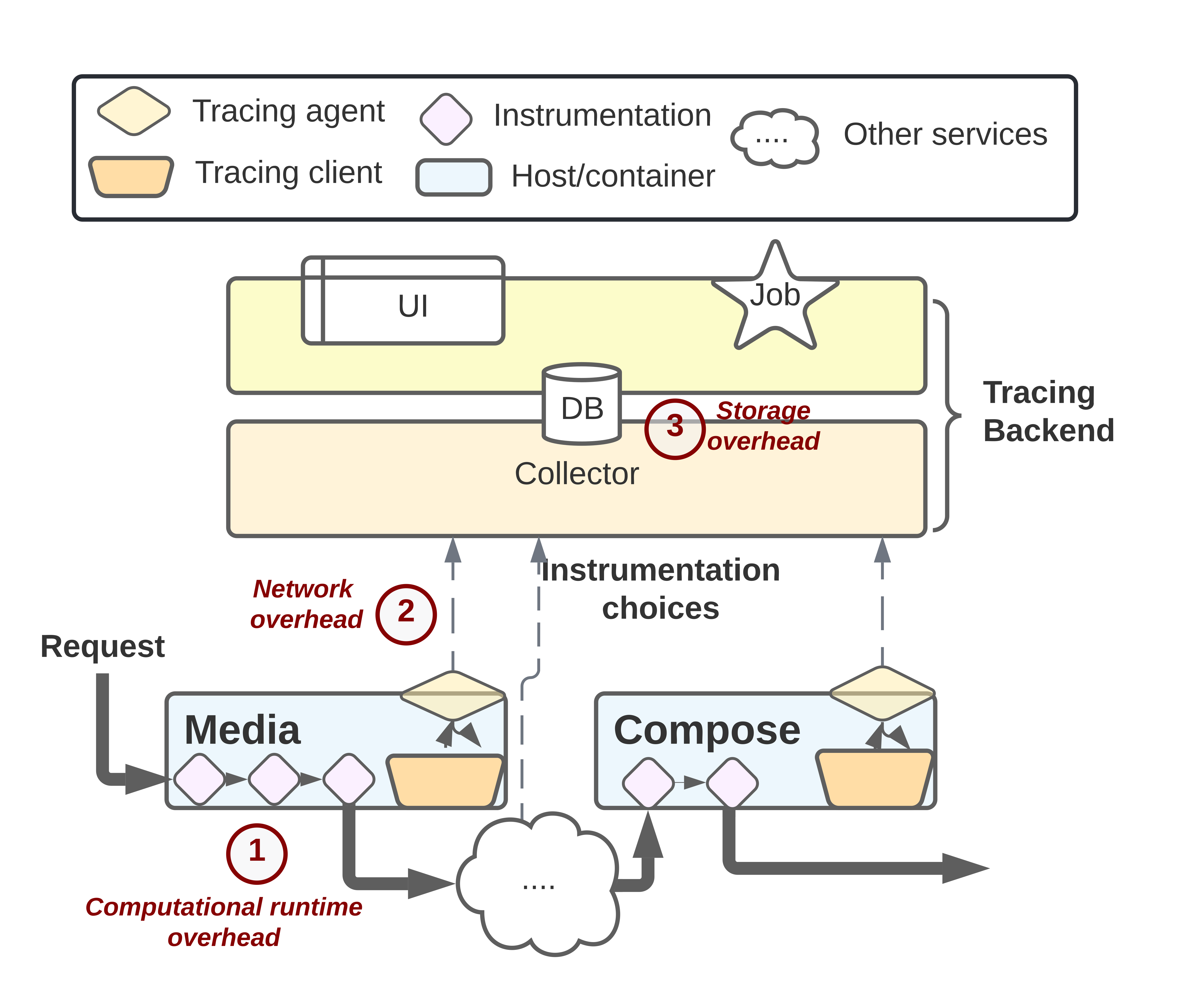}
\centering
\caption{An overview of the distributed tracing architecture. The bottom shows a simplified version of the Social network application, instrumented with tracing. The top shows the tracing backend. Tracing overheads are demonstrated with circled numbers.}
\label{fig:tracingframework}
\end{figure}

\subsection{Distributed tracing overview}
\label{sec:tracingoverview}

Figure \ref{fig:tracingframework} shows how distributed tracing frameworks operate in general \cite{canopy,jaeger}.
An instrumented service (e.g., Media in Fig. \ref{fig:tracingframework}) creates a \textit{span} when receiving a new request and attaches context information (e.g., traceId and spanId) to outgoing requests. 
A span represents a logical unit of work with an operation name, start time, and duration. Spans are nested and ordered to model causal relationships (i.e., parent/child relationships represent caller/callee). 
The tracing backend collects, orders, and stitches span records with the same traceId to create end-to-end traces. 
A final trace is an execution path of a request through the system \cite{vaif}.

\begin{figure}[t]
    \includegraphics[trim={0.18cm 0.1cm 0.11cm 0.28cm},clip,width=0.8\columnwidth]{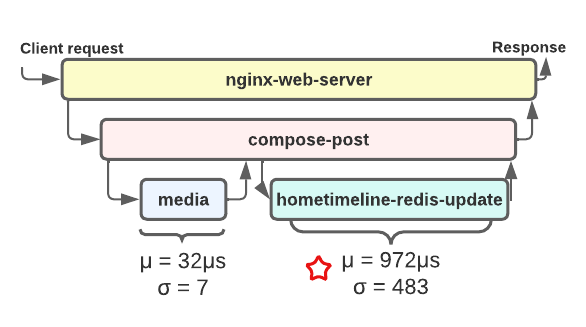}
    \centering
    \caption{A simplified trace from Social network. Analysis on $\mu$ and $\sigma$ of span latency helps localize a performance issue.}
    \label{fig:redisproblem}
    \end{figure}

The causal ordering of spans enables composing a narrative of a request's execution, making it easier to pinpoint issues.
The following example illustrates how distributed tracing helps diagnose a performance issue we found using \tool~ (\cref{sec:cases}).
Fig. \ref{fig:redisproblem} demonstrates a simplified trace, along with various statistics of operation durations ($\mu$ and $\sigma$) from the Social network application. 
The root cause of the issue is an inefficient implementation that sends one request for each key to Redis instead of querying with multiple keys \cite{redisplus,redis-sn}.
Analysis via distributed tracing reveals that the Redis update operation shows high variance and significantly contributes to response time. 

Distributed tracing helps pinpoint the code regions that initiate performance problems \cite{tprof,sambasivan2011diagnosing}.
However, there are challenges with efficient and practical use of distributed tracing in production.
Developers wish to comprehensively trace the source code
to easily diagnose new problems \cite{vaif,log2}; however, they struggle with the cost vs. utility trade-off \cite{log20}. 
Reducing latency is one of the primary concerns that lead developers to use distributed tracing. 
However, overheads of distributed tracing also contribute to latency, as shown in Fig. \ref{fig:tracingmotivation}.
Understanding this impact is vital in choosing the right granularity for tracing. 
We briefly describe overheads with respect to Fig. \ref{fig:tracingframework} (marked with circled numbers). 
\circled{1} Spans are created on the critical path of requests.
This step incurs {\em computational runtime overhead}.
\circled{2} Tracing agents listen for spans from the application, receive, and route them to the tracing backends. This step incurs {\em network overheads}.
\circled{3} Tracing backends persist the trace data, where they can be later queried. This step incurs {\em storage and compute overheads}.


\subsection{Related work}
\noindent\textbf{Distributed tracing \& diagnostics. }
Past research has shown a variety of use cases of distributed tracing in cloud \cite{jackpot,sambasivan2011diagnosing,delimitsage,delimitseer,tprof,canopy,ardelean2018performance,chow2014mystery,xtrace,sifter,universal-mace,dapper,firm,zeno,Parker,cripscp}.
Zeno \cite{zeno} uses traces to infer off-path root causes such as resource contentions.
SAGE \cite{delimitsage} is a machine learning-based system that leverages tracing to identify culprit services responsible for QoS violations.
These state-of-the-art systems aim to facilitate rich use cases but are not geared towards resolving the cost vs. utility of the spans trade-off.

\noindent\textbf{Request-based sampling. }
To help with various overheads of distributed tracing, existing techniques include head- and tail-based request sampling~\cite{jaeger,dapper,opentelemetry,sifter}.
Head-based method decides whether to trace a request uniformly at random at the beginning of a request, circumventing computational, network, and storage overheads (\circled{1}, \circled{2}, and \circled{3} in Fig. \ref{fig:tracingframework}).
On the other hand, tail-based methods capture traces for all requests and later decide whether to persist a trace, circumventing only storage overheads (\circled{3} in Fig. \ref{fig:tracingframework}).
For example, Sifter \cite{sifter} biases sampling decisions towards outlier traces with respect to their frequency.
While effective at reducing various overheads, these techniques are orthogonal to the cost vs. utility of the spans trade-off.
That is, they don't control the instrumentation at the granularity of spans while tracing; 
thus, resultant traces include a large portion of extraneous spans that constitute the majority of costs.
When diagnosing the current performance variation, only a few spans are useful, as a large portion of them are extraneous, constituting the majority of cost~\cite{vaif,log2,pivot}. For example, the Fig.~\ref{fig:redisproblem} shows that 16\% of spans constitute more than 80\% of the total variation.

\noindent\textbf{Dynamic instrumentation. } 
Some techniques allow inserting instrumentation at runtime in almost arbitrary locations in applications \cite{pivot}.
These techniques provide crucial flexibility during performance debugging. 
However, manual and iterative exploration of potential instrumentation locations in source code is a labor-intensive process, incurring prolonged diagnosis times~\cite{tracinginpractice}.
Unlike dynamic instrumentation, \tool~does not rely on end users to manually explore data and identify problems but automates this process.

\noindent\textbf{Automated control of instrumentation. }
To address cost vs. utility trade-off, researchers have proposed automated techniques that selectively enable instrumentation~\cite{vaif,log2,
log20}.
However, most focus on correctness (or failures), not performance variations. 
For example, Log20 \cite{log20} enables log points to differentiate request workflows, aiming to identify faulty executions.
Such methods are not sufficient for performance diagnosis because additional instrumentation may be needed to identify where on a unique workflow a performance problem lies.
In contrast, \tool~focuses on identifying areas of the code that lead to requests’ execution to be slow. 
For example, inefficient implementation of a Redis update might not break functionality, but can cause a significant slowdown (Fig. \ref{fig:redisproblem}).
The systems closest to our work are VAIF and Log\textsuperscript{2}, tailored for performance variations. 
We compare \tool~with these systems in \cref{sec:experiments}.

%% file: design.tex
\begin{figure}[t]
  \includegraphics[width=1.0\columnwidth]{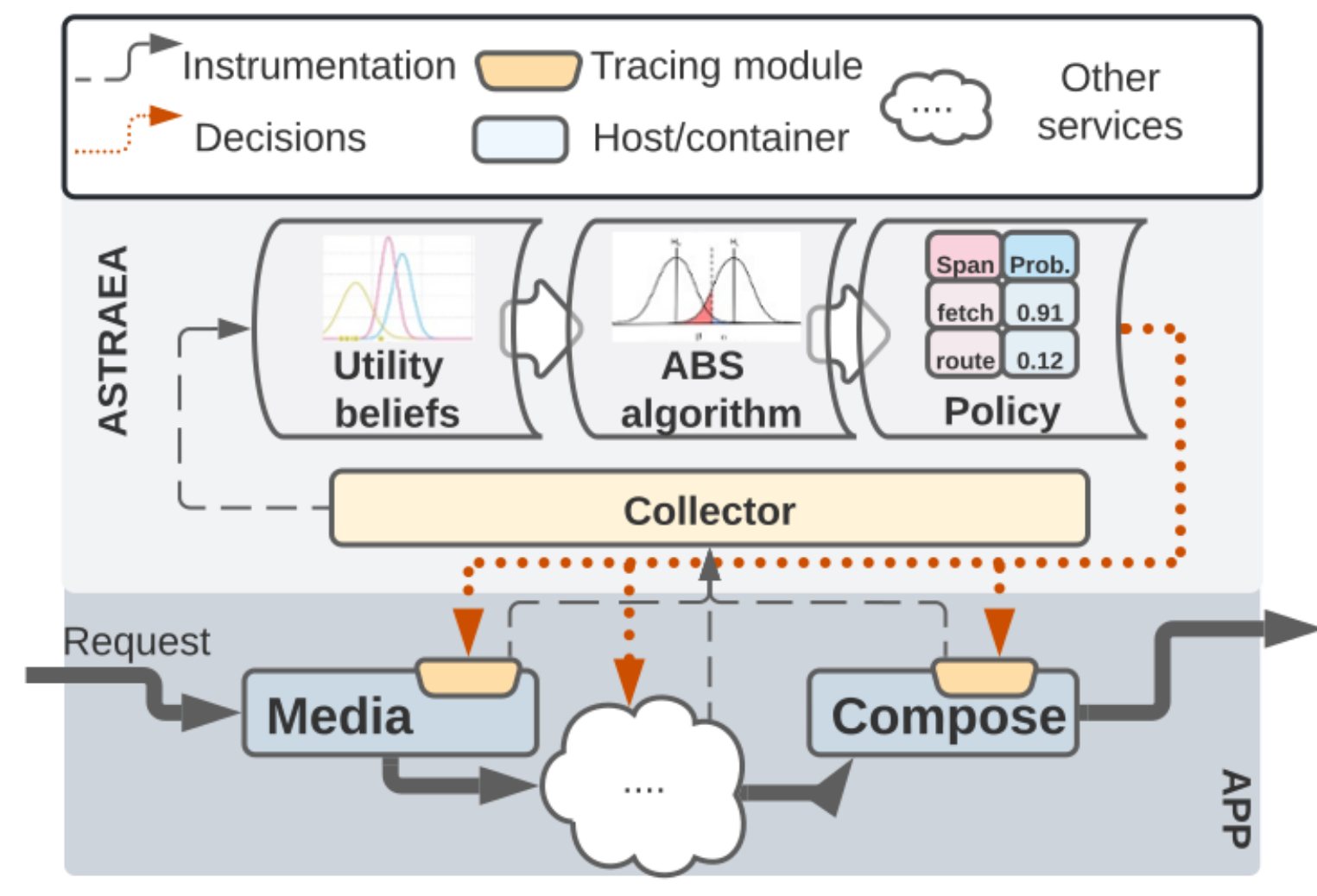}
  \centering
  \caption{\tool~  design. The bottom displays an instrumented application, and the top features \tool~ components guiding tracing to rewarding spans.}
  \label{fig:astraea}
\end{figure}

While an ideal tracing solution would focus effort on parts of an application that are important for performance diagnosis, this information is unknown \textit{a priori}, hence the need for learning at runtime. \tool~is an online, probabilistic tracing system designed to combat this challenge. 
Fig \ref{fig:astraea} shows \tool~components that implement the logic of effectively learning the parts of an application that are needed for diagnosis (exploration) and confidently steering tracing toward the most rewarding instrumentation choices (exploitation). 
\tool~works in a continuous loop.
As the application receives user traffic, trace observations are collected and stored in a database. 
\tool~ periodically queries the tracing database to build Bayesian belief distributions to estimate span utilities. 
Utility represents the usefulness of a span for performance diagnosis (\cref{sec:utility}); as more data is available, the belief distributions converge to the true values of the utility (\cref{sec:belief}). 
Using accumulated beliefs,
\tool~ populates a policy, which is a dictionary that maps span identifiers to their sampling probabilities. Fig. \ref{fig:astraea} shows a snippet of a sample policy, where \textit{fetch} is the span and its sampling probability is $0.91$.
\tool~ continuously builds and updates this policy to steer tracing toward vital spans, using its
Approximate Bayesian Sampling algorithm considering exploration vs. exploitation trade-off (\cref{sec:bandit}).
The tracing modules on the bottom are modified to deliver a lightweight mechanism of enabling and disabling spans centrally controlled by \tool~ (\cref{sec:impl}).
The policy is periodically issued to the tracing module in the application via \tool.
The tracing module uses this policy to generate the spans according to their respective probability.


\noindent\textbf{Output.}
\tool~is intended to help developers diagnose unanticipated performance variations by dynamically controlling the instrumentation in running systems concerning the cost vs. utility of spans trade-off.
The main output from \tool~ is, in fact, distributed traces; however, the data contained in the traces is the set of vital spans that are most able to explain performance variations.
Second, \tool~provides users a query interface that reports the ranking of spans based on their utilities, top-k problematic spans, confidence levels (e.g., the probability that a span is vital), and significantly correlated tags (e.g., service.version) within traces
to help developers interpret the results regarding localized code regions (\cref{sec:cases}). 

\noindent\textbf{Design principles.}
\tool~ adheres to the following principles to address practicality, accuracy, and efficiency requirements (\cref{sec:intro}).
\begin{itemize}[leftmargin=*]

  \item \textbf{Statistical rigor:} 
  While \tool~doesn't offer explicit accuracy guarantees, our method diverges from a greedy strategy of consistently opting for the highest variance span. Rather, \tool~ constructs Bayesian belief distributions, guiding decisions based on statistical confidence. Consequently, it demonstrates superior accuracy in pinpointing performance issues, as evidenced by the experimental results (\cref{sec:experiments}).
  \item \textbf{Adaptive:} 
  Binary decisions on enabling (or disabling) spans prevent existing tools from adapting to changing sources of variations.
  \tool's probabilistic approach enables keeping pace with the changes on-the-fly (\cref{sec:onlinelearning}).
  \item \textbf{Low-overhead \& scalable:} 
  \tool's low-dimensional Bayesian approach enables space- and time-efficient computations to decide whether to create spans based on accumulated beliefs, which helps avoid computational, network, and storage overheads (\circled{1}, \circled{2}, and \circled{3}).
  Because \tool~maps large trace data to compact Bayesian representation, it circumvents storing large chunks of trace and can scale well with production workloads (\cref{sec:production}).
  \item \textbf{Online:} 
  Training the entire tracing solution is cumbersome, exacerbated by retraining due to frequent code updates. 
  \tool~ embodies an online learning framework to eliminate offline phases.
\end{itemize}

\subsection{Span utility}
\label{sec:utility}
We now examine how \tool~decomposes latency contributions of spans and present the span utility measure, which assesses the effectiveness of spans in performance diagnosis.

\begin{figure}[tb]
  \includegraphics[trim={0.28cm 0.1cm 0 0.28cm},clip,width=0.8\columnwidth]{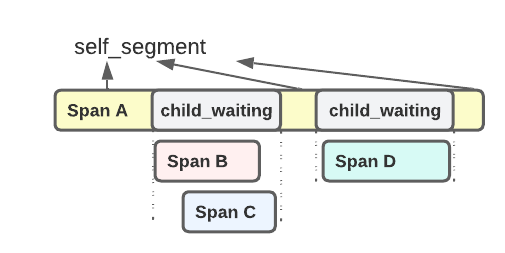}
  \centering
  \caption{The latency contribution of a leaf span (B, C, and D) corresponds to its processing duration. In the case of non-leaf spans, \tool~ isolates the self\_segment, representing the duration when an operation is not awaiting the completion of a child operation.}
  \label{fig:self}
  \end{figure}

\noindent\textbf{Latency decomposition.} Fig. \ref{fig:self} shows the latency decomposition of spans. 
Trace includes a root span that corresponds to the client's request to the web server (span A) that calls various operations in parallel (span B and C) and sequentially (span D) after receiving responses from preceding operations. 
The latency contribution of a leaf span (e.g., span D) is determined by the processing duration of itself.
On the other hand, a non-leaf span with one or more children (e.g., span A) is further decomposed into segments (i.e., self\_segment and child\_waiting). 
\tool~ relies on the self\_segment that represents the amount of time spent by span itself.

\noindent\textbf{Utility.}
\tool~ defines the \textit{utility} as something that represents how much a span contributes to performance diagnosis. The utility serves the following two purposes. 
First, it is considered a \textit{reward function} in \tool's online learning framework to identify and focus on spans with the highest utility \cite{Sutton1998}.
Second, utilities can also be used to serve insights into developers' diagnosis efforts in terms of ranking and comparison of different spans.
We study various ways to measure the utility of a span through performance anomaly injection experiments on three cloud applications (Social network, Media, and Train ticket) and 
problems include random delays and resource contentions (\cref{sec:apparatus}).
\begin{figure}[tb]
\includegraphics[width=0.8\columnwidth]{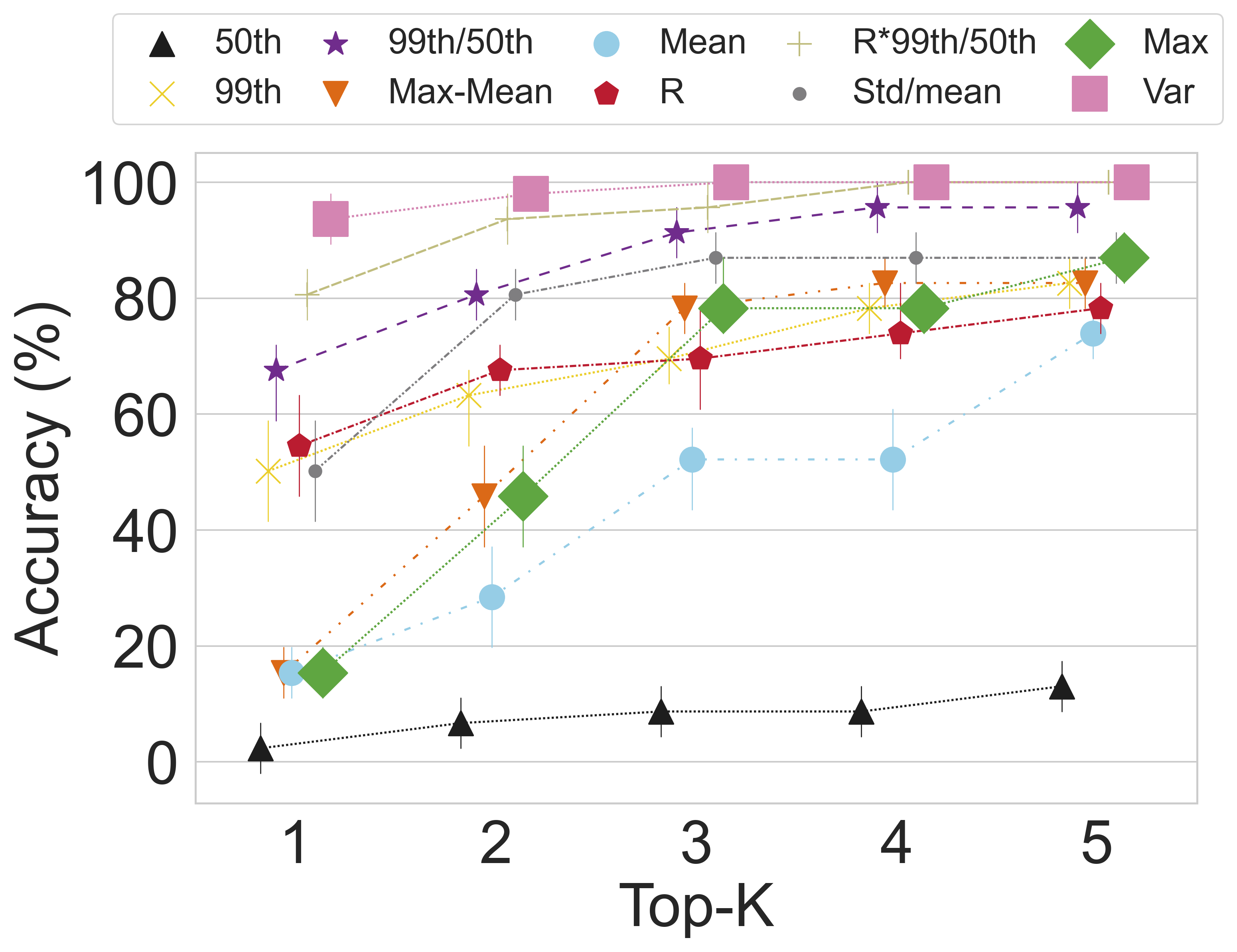}
\centering
\caption{Various statistical measures for span utilities. Figure evaluates accuracy in terms of whether top-k spans (with the maximum sampling
probabilities in \tool~ with given utility) capture the faulty spans (correspond to the location of injected problem). }
\label{fig:utility}
\end{figure}
Figure \ref{fig:utility} shows the top-K accuracy (i.e., whether the correct span is among the top-K rankings) per statistical measure.
We find that the latency variance yields the most favorable outcomes, aligning with recent research \cite{vaif, firm}, which consequently becomes the default utility measure in \tool.
Beyond preset utility measures, \tool~allows developers to define and use their own customized measures for specific scenarios.


\noindent\textbf{Tags analysis.}
\tool~records each tag pair in spans and constructs a data structure, where features represent tag values (categorical ones encoded using sklearn LabelEncoder), and the target variable is end-to-end latency. \tool~applies Pearson correlation analysis to measure the strength of relationships between tags and latency. 
Analyzing tags is crucial for performance diagnosis; for instance, we illustrate how the service.version tag can reveal issues with newly deployed code versions (\cref{sec:cases}).


\input{bayesian.tex}

\subsection{Implementation}
\label{sec:impl}

Our \tool~ prototype was developed in Python (2.5K LOC), utilizing popular data science libraries such as NumPy, SciPy, and Pandas. 
Our implementation involved several key components of \tool, including a module that periodically utilizes Jaeger APIs to fetch batch of trace data and derives sampling policy and a module that applies sampling decisions to the instrumented applications.

During each cycle, \tool~ fetches a batch of trace data and converts them to a low-dimensional Bayesian format. To be specific, we assign a belief distribution to each span with alpha and beta parameters initialized to 1, ensuring that the prior beliefs are uniformly random. Based on the span utilities acquired from the batch of trace data, belief distributions are updated. 
The probability that a span belongs to the vital set is approximated using Monte Carlo sampling, which is the most computationally intensive task. 
To enable fast and efficient computations for this step, we utilize NumPy, which implements statistical sampling functionality in C.
\tool~ periodically applies span sampling policies to the tracing modules in the application layer. These policies map span identifiers to their corresponding sampling probabilities. We use a combination of service name, operation name, and URL to identify individual spans, addressing instances of identical operation names.
Utilizing \tool~ in cloud applications is a straightforward process. These applications only need an existing distributed tracing framework to enable the required support. 
We developed two prototype tracing modules in Java (for Train ticket) and C++ (for Social network and Media) languages to utilize \tool.
Our implementation introduced a low-priority background thread that periodically receives sampling policy from \tool~ and loads it into memory as a dictionary.
To store the policy data produced by \tool, we utilize Docker volumes that are accessed by the modified Jaeger tracing module living in Docker containers. Our modifications were kept to a minimum, with the tracing module only fetching span policy data from the volume at periodic intervals (e.g., every 5 seconds).
Our policy data structure includes per-service span identifiers and their sampling probabilities, with low memory overhead (a few kilobytes).

In Jaeger C++, we have only made changes to the function responsible for starting spans. Our changes include querying the in-memory policy to acquire the sampling probability of the intended span and determining whether to create the span based on that probability.
In Jaeger Java, we needed to make changes in opentracing-spring-jaeger library that provides automated instrumentation for Java Web Servlet applications.
Specifically, we modified the servlet filter that is used to create spans upon receiving requests in Spring stack. 
Similar to C++ implementation, we implement the sampling decisions based on the probabilities fetched periodically from \tool.
The modifications needed to deliver selective enabling or disabling of spans by \tool~ were minor, totaling less than 200 lines of code.
In our implementations, if a span is not sampled, the function to start spans does not create new spans but returns, which frees up overheads such as computational costs (e.g., allocating a new object, attaching a reference, reading from a performance timer, and updating the thread-local state), network costs (e.g., emitting the span), and storage costs (e.g., persisting the span record) for the disabled spans.
In contrast to the alternatives of tail-based sampling (Log\textsuperscript{2}) or conditional checks using filesystem (VAIF), our implementation uses an in-memory sampling policy to make sampling decisions based on accumulated beliefs before creating spans, resulting in significantly lower overhead (as shown in the evaluation).


The main output of \tool~ is distributed traces that contain vital spans for explaining performance variations. When developers use the tracing UI to address issues, they will only see crucial spans that explain the current variations. Additionally, \tool~ reports the ranking of spans based on their usefulness, the most problematic spans, confidence levels (such as the probability that a span is vital), and tags that are correlated with latency. 

%% file: bayesian.tex
\subsection{Online learning}
\label{sec:onlinelearning}
\tool's online learning framework encapsulates our
Bayesian online learning formulation and multi-armed bandit-based
sampling algorithm (ABS). 
The former builds belief
distributions to progressively learn span utilities, and the latter adjusts span sampling probabilities based on the belief distributions.

\noindent\textbf{Belief distributions.}
\label{sec:belief}
\tool~ associates a belief distribution with every
span and updates them periodically based on the new batch of tracing data. The batched update is
a practical necessity, as traces are not available instantaneously but often with delays \cite{iter8,jaeger}.

We present the mathematical intuition behind belief updates using the following scenario. 
Suppose we have a coin with $p$ ($=0.7$) as the probability of landing a head.
In Bayesian inference, $p$ is a random variable and can be modeled using beta belief distributions. Beta distribution, $Beta(\alpha,\beta)$, describes how likely $p$ can take on each value between 0 and 1. 
$\alpha$ and $\beta$ parameters can be thought of as the number of successes and failures, respectively. $Beta(1,1)$ is a reasonable \textit{prior} when we have no a priori information about $p$ as it distributes $p$ uniformly in [0, 1].
If we toss the coin 100 times and observe 70 heads in the sample,
 we can use Bayes' theorem to update our belief for $p$ (i.e., $\alpha$ $\gets$ $\alpha$ + \# of heads (vice versa for $\beta$)) to obtain a \textit{posterior distribution} for $p$, which is now $Beta(1 + 70, 1 + 30)$. 
This is an instance of the classic beta belief update model.


Drawing from the coin example, \tool~ models span utilities in the form of beta distributions. 
Utility measures such as variance of latency may not be 0/1 random variables; thus, \tool~ normalizes the utility observations with respect to the maximum value observed in samples.
Formally, 
let $max(\overline{h(s)})$ represents the largest value of the sample utility function $\overline{h(s)}$.
The beta distribution for span $s$, $Beta(\alpha_{s}, \beta_{s})$, captures the uncertainty about the \textit{normalized expectation} $\frac{\mathbb{E}[h(s)]}{max(h(s))}$. 
This distribution is initialized to uninformative prior, $Beta(1,1)$. \tool~ then updates these distributions periodically at each iteration based on the new batch of trace observations. 
Let $\overline{h_{e}(s)}$ represents the sample utility for span $s$ at epoch $e$, and the beta belief updates for each span are illustrated as follows:

\begin{gather}
\alpha_{e, s} \gets (1-\lambda) * \alpha_{e-1, s}  + \lambda* \frac{\overline{h_{e}(s)}}{max(\overline{h(s)})} \label{eq:alpha} 
\\
\beta_{e, s} \gets (1-\lambda) * \beta_{e-1, s} + \lambda* \frac{max(\overline{h(s)}) - \overline{h_{e}(s)}}{max(\overline{h(s)})} 
    \label{eq:beta} 
\end{gather}
where $\lambda$ represents exponentially weighted moving average (explained below). We choose the Bayesian beliefs due to its simplicity and the following considerations. 
First, it provides a low-dimensional data structure, enabling space- and time-efficient computation (\cref{sec:experiments}).
Second, the posterior mean of the beta distribution approaches true expectation almost surely as with more data, provided by the law of large numbers \cite{degroot:book}. Similarly, the variance of the distribution, $\frac{\alpha\beta}{(\alpha + \beta)^2(\alpha + \beta + 1)}$, goes to 0 as with more samples. 
Third, \tool~ can easily incorporate new span utility distributions (e.g., due to a code change). 
Initially, the belief distributions for the new spans
will exhibit higher variance; however, they will converge to
their true expectations as with more requests.

\textit{\textbf{Non-stationary distributions.}}
It is common to find that span utility distributions drift over time (e.g., due to changing performance problems). To address this, \tool~ employs $\epsilon$-exploration strategy that prevents span sampling probabilities from reaching zero, ensuring observation of new samples with minimum probability of $\epsilon$ (default 5\%).
In \cref{sec:sensitivity}, we experiment with various values of $\epsilon$ to provide insights on this parameter.


\begin{figure}[tb]
    \includegraphics[width=0.8\columnwidth]{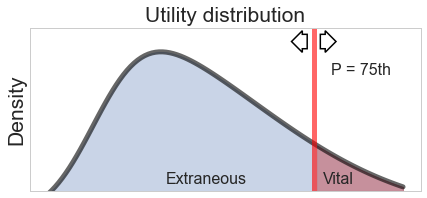}
    \centering
    \caption{Percentile-based threshold determines the vital set of spans that contribute most to performance variation.}
    \label{fig:percentilethreshold}
    \end{figure}

\noindent\textbf{Approximate Bayesian sampling.}
\label{sec:bandit}
\textit{Given the span observations until now, how can we quantify the extent to which a span is worth sampling?}
Fig. \ref{fig:percentilethreshold} shows a representative span utility distribution, similar to what we observed (right-skewed) in traces from benchmark applications and the production system (\cref{sec:experiments}). 
\tool~ employs a percentile-based threshold (denoted by \textit{P}, default $75^{th}$) to determine whether a span belongs to the \textit{vital set} that contributes most to performance variation.
We choose the percentile-based threshold for the following reasons. 
First, the percentile is an intuitive measure that depicts where a span stands relative to others. 
\tool~ focuses tracing effort on spans that fall above the \textit{P} (i.e., vital set).
Second, the percentile serves as a control knob, modeling the developers' preferences. 
By setting \textit{P} to higher values (e.g., $90^{th}$), developers can increase the focus on the tail portion and enjoy a higher reduction in trace sizes (\cref{sec:sensitivity}). 

The probability that a span belongs to the vital set can be approximated using random sampling, referred to as Monte Carlo methods \cite{degroot:book}.
Recall that $Beta(\alpha_{s}, \beta_{s})$ represents the posterior belief distribution for the normalized utility ($\frac{\mathbb{E}[h(s)]}{max(h(s))}$). 
\tool~ samples multiple times from this distribution, and the collected data is used to approximate the desired quantity. Given the law of large numbers \cite{degroot:book}, larger number of random samples enable a more accurate approximation (e.g., $1\mathrm{e}{+6}$ in \tool).

Given the posterior probabilities, \tool~ now faces a fundamental trade-off between exploration (of span utilities) vs. exploitation (confidently steering tracing toward vital spans). 
Exploration-exploitation trade-off is best exemplified by the multi-armed bandit problem \cite{thompson:sampling,iter8}. 
Drawing from the bandit problem, our ABS algorithm chooses the spans with probability matching decision strategy \cite{thompson:sampling}.
To illustrate, suppose that span A belongs to vital set 90\% of the samples and span B with 10\%.
A greedy strategy is to always enable A and disable B, hence can forgo an opportunity to learn about span B.
In contrast, ABS samples spans A and B with probabilities equal to 0.9 and 0.1, respectively.
As such, ABS explores to resolve uncertainty to help identify the optimal spans.

More formally, \tool~ uses the Monte Carlo procedure to create a sample utility approximation matrix with each row containing a single utility sample for each span.
Given the matrix, 
ABS picks the spans that exceed the given percentile \textit{P} of utility as per row of the matrix and marks them as candidates. 
In the final sampling policy, the sampling probability of a span equals the fraction of rows in which the span emerged as the candidate. 
Since the posterior probability of suboptimal spans converges to 0 over time (\cref{sec:belief}), ABS progressively shifts tracing towards the optimal spans as desired. Our algorithm is inspired by the classic Bayesian algorithms designed for such explore vs. exploit problems \cite{thompson:sampling}. These algorithms provide various benefits over classical multivariate hypothesis tests and have solid theoretical guarantees.

\textit{\textbf{$\epsilon$-ABS. }} \tool~employs \textit{$\epsilon$-exploration} strategy to account for non-stationary behavior. So, final sampling policy computed by \tool~ is equal to maximum of the probability that ABS derives and $\epsilon$.

%% file: eval.tex
We evaluate \tool~  on three distributed cloud applications (Social network, Train ticket, and Media) and production traces acquired from a large-scale Internet company.
Our experiments encompass:
(i) validation of \tool~  in accurately steering tracing toward injected performance issues and reduction in trace sizes compared to unmodified Vanilla Jaeger tracing (all spans enabled);
(ii) comparative analysis with two state-of-the-art baselines, VAIF and Log\textsuperscript{2};
(iii) sensitivity analysis of \tool~  to various parameters; 
(iv) scalability analysis to production traces.

\subsection{Experimental apparatus}
\label{sec:apparatus}
\noindent\textbf{Benchmark applications \& datasets.}
We use three distributed cloud applications and a trace dataset acquired from a large-scale Internet application to evaluate \tool.
\textit{Social network} is a broadcast-style social network, implemented with 36 microservices, including databases, caches, application logic, and a frontend Nginx web server, communicating with each other via Thrift RPCs. 
The application is instrumented with the Jaeger-CPP client~\cite{jaegercpp}.
\textit{Train ticket} is a publicly available booking system, implemented with 41 microservices that communicate via REST APIs.
The application uses Opentracing java library \cite{opentracingspring} that automatically instruments the Spring stack by recording each web request/response via Jaeger-Java client~\cite{jaegerjava}.
\textit{Media} implements an end-to-end service for browsing movie information, reviewing, renting, and streaming movies. The Media application consists of 38 services, and they are instrumented with Jaeger-CPP client.
\textit{Large-scale Internet application dataset} comprises $\sim$20K obfuscated traces from a production system, using which
we demonstrate \tool's scalability to large-scale systems (\cref{sec:production}). 

\vspace{0.1in}
\noindent\textbf{Comparison baselines.}
We compare our proposed system to two automated techniques that selectively enable the instrumentation required for performance diagnosis.
\textit{Log\textsuperscript{2}~\cite{log2}} is an automated logging mechanism to decide `whether to log' in such a way that the logging overhead is constrained within the budget while maximizing logging effectiveness.
\textit{VAIF~\cite{vaif}} is an automated instrumentation framework that integrates distributed tracing with control logic to dynamically enable the necessary instrumentation for identifying performance issues. During an offline profiling phase, VAIF maps out all possible execution paths of requests, which are referred to as the constructed search space.

\vspace{0.1in}
\noindent\textbf{Infrastructure setup.}
We performed experiments on a Cloudlab cluster comprising four nodes, all running Ubuntu 18.04 with Linux 4.15, in order to assess the effectiveness of our proposed system.
Each node had 8 CPU cores (Intel(R) Xeon(R) E5-2640), 64 GB of memory, and a 1 Gbps network connection. 
We deployed the aforementioned applications separately using a YAML configuration file that defines a set of services and their dependencies (Docker version 23.0.2).
We used hey \cite{heyworkload} and wrk2 \cite{wrk2workload} workload generators, to send requests to applications with various concurrent workers (between 5-25) and req/s (between 20-100).
We conducted a study to compare our approach with baselines using three different applications.
We introduced various types of performance degradations to randomly chosen operations or services~\cite{firm,tailatscale}.
We introduced random delays with sleep functions placed in source code of applications. The intensity and probability of delays were randomly sampled from normal distributions with various means (e.g., 1 ms to 10 ms). Additionally, we used Pumba~\cite{pumba}, a chaos testing command-line tool for Docker containers, that uses CPU and memory stressors of stress-ng\cite{stressng}, resulting in delays in the upstream operations of congested containers. As an oracle, we determined the faulty spans that captured injected anomalies and experimented to determine if our approach could capture those spans.

\subsection{\tool~  validation}
\label{sec:accuracy}
We first validate \tool's ability to correctly localize problematic operations and steer tracing toward them. 

\begin{figure*}[t!bhp]%
    \centering
    \subfloat[Sampling of faulty span through time\label{fig:probs}]{{\includegraphics[width=0.7\columnwidth]{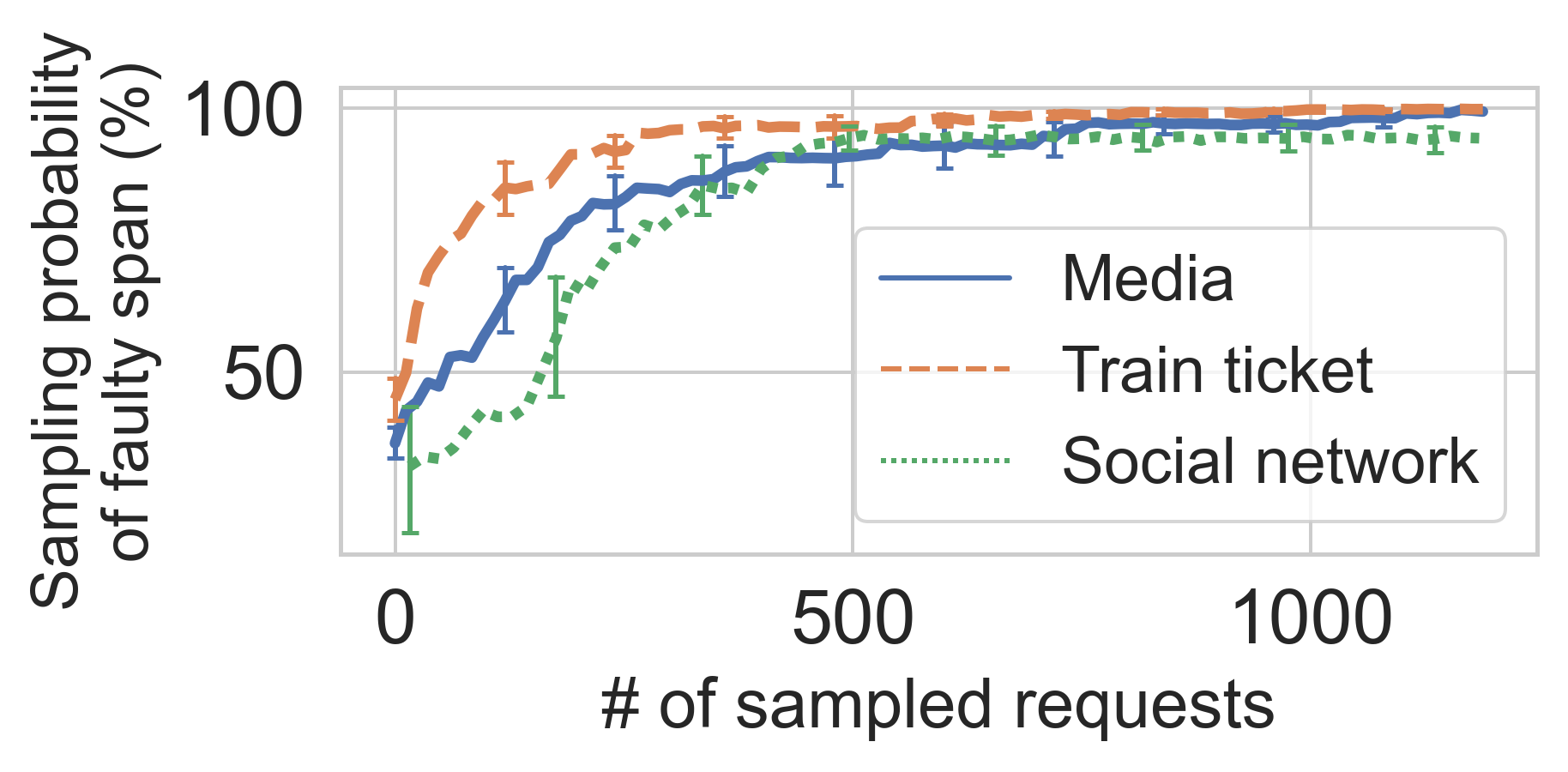} }}
    \subfloat[Accuracy of top-k spans (w/ highest sampling prob.)\label{fig:acc}]{{\includegraphics[width=0.7\columnwidth]{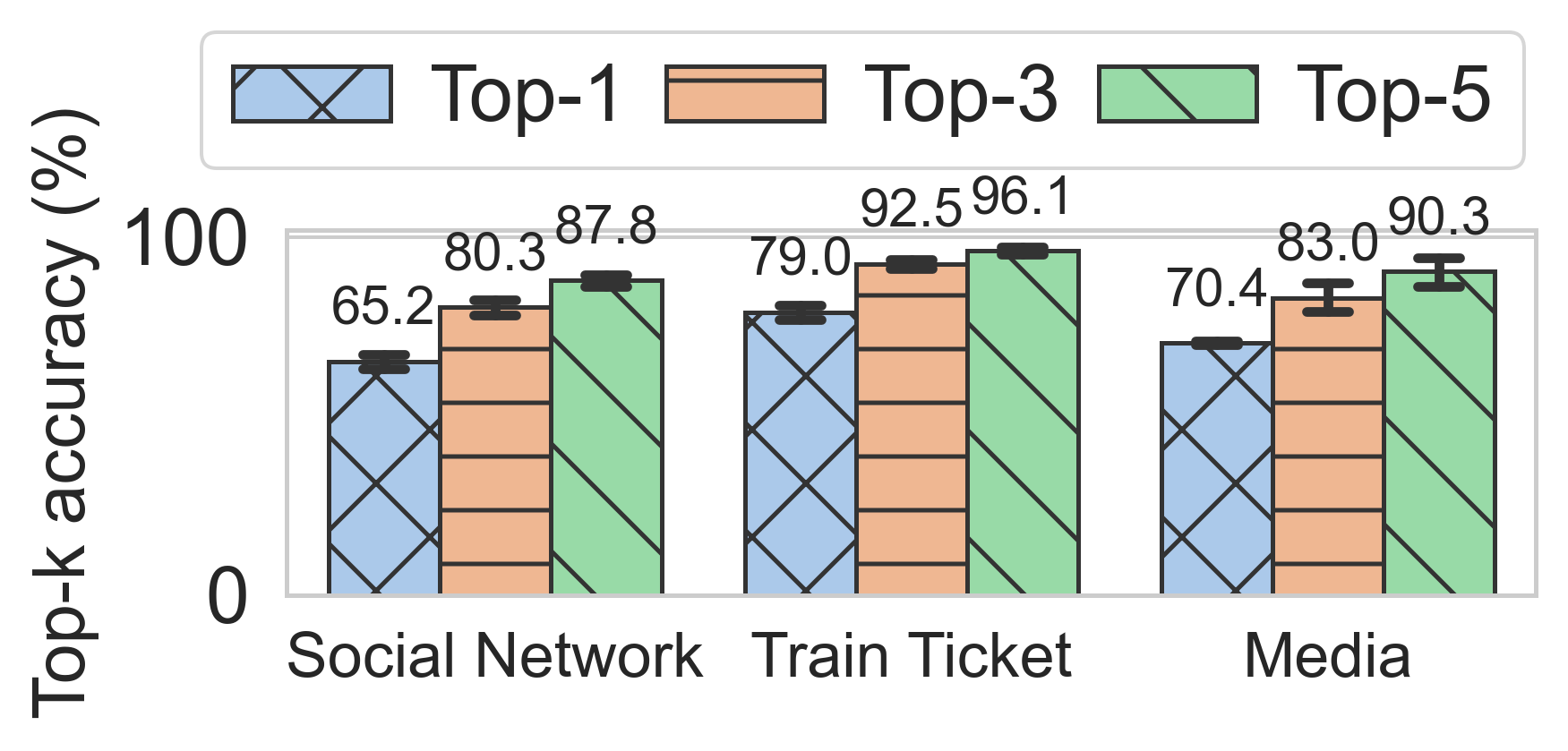} }}
    \subfloat[Trace size comparison with Vanilla Jaeger\label{fig:savings}]{{\includegraphics[width=0.6\columnwidth]{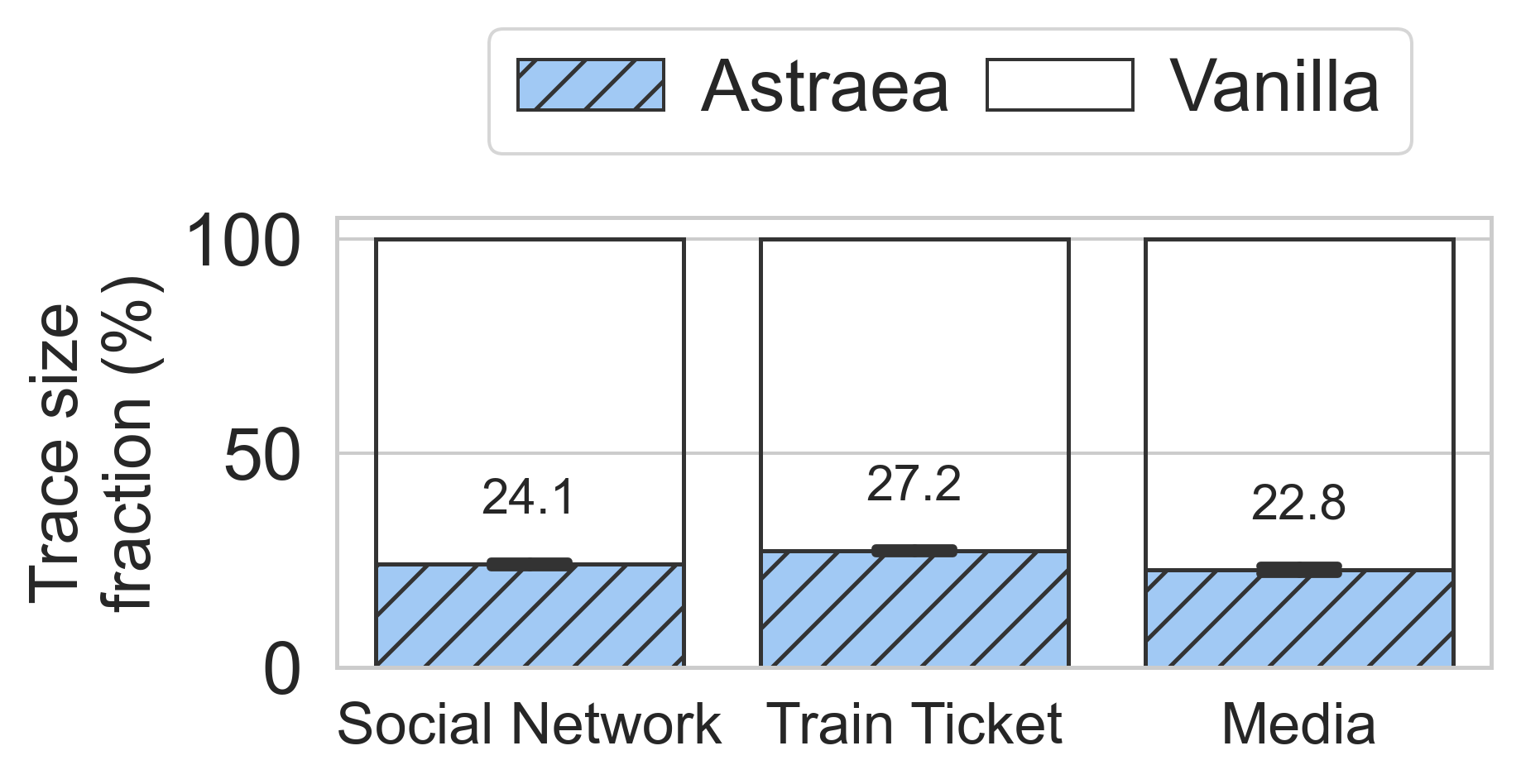} }}
    \caption{\tool's accuracy and reduction in trace sizes. \tool~  quickly maximizes the sampling probability of faulty spans (where problem is injected) to above 90\% confidence level within $\sim$300 samples on average, enabling only $\sim$24\% of all available spans cumulatively.}%
    \label{fig:astraeaacc}%
\end{figure*}

Fig. \ref{fig:probs} reports the sampling probability of faulty spans, which capture problematic operations from 20 different runs per application.
\tool~  quickly maximizes the sampling probability of faulty spans in all applications (above 90\% within $\sim$300 samples on average), achieving 91\% cumulative coverage. Coverage, is the average sampling probability of the faulty span for the experiment.

Fig. \ref{fig:acc} presents whether top-k spans (with the maximum sampling probabilities in \tool) capture the faulty spans.
\tool~  achieves 92\% Top-5 accuracy on average. 
Accuracy is slightly higher for Train ticket as
traces include a larger number of repeated spans and sequential executions, enabling \tool~  to learn faster.
Fig. \ref{fig:savings} demonstrates the fraction of traces that \tool~  uses with respect to the Vanilla Jaeger implementation, where all spans are enabled.
We find that \tool~  only enables 24.7\% of instrumentation cumulatively while accurately tracing problematic operations.
Trace sizes in Train ticket are slightly larger due to a larger number of repeated spans. 

\subsection{Comparative analysis}
\label{sec:comparison}
We compare \tool~  with VAIF and Log\textsuperscript{2}. 
We run VAIF with default parameters \cite{vaif}.
Log\textsuperscript{2} requires a budget parameter \cite{log2}, which we set equal to the average number of spans enabled by \tool.

\noindent\textbf{Accuracy.}
\begin{figure*}[t!bhp]%
    \centering
    \subfloat[Coverage of faulty spans in traces\label{fig:compcover}]{{\includegraphics[width=0.66\columnwidth]{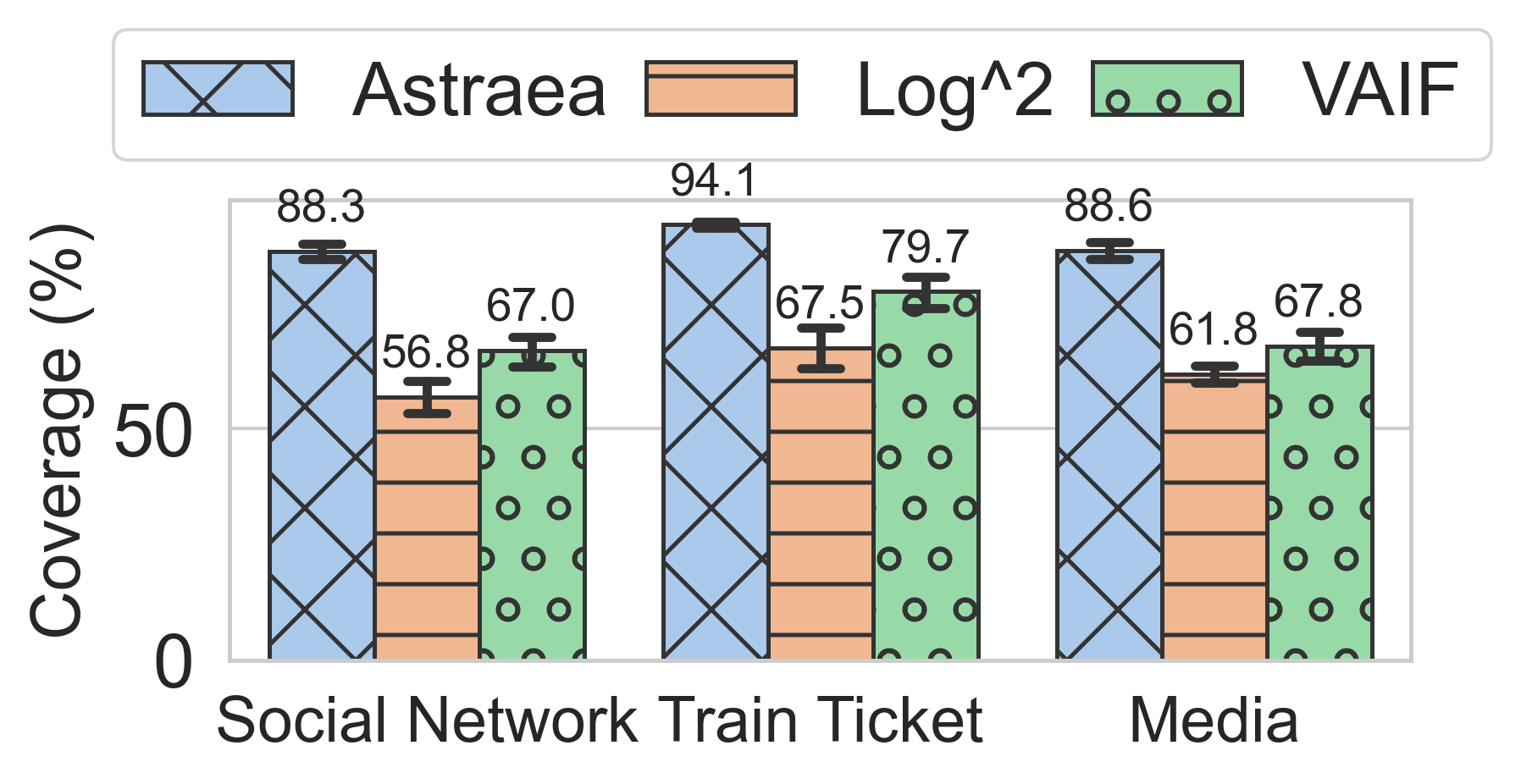} }}
    \subfloat[Trace size comparison\label{fig:compsizes}]{{\includegraphics[width=0.66\columnwidth]{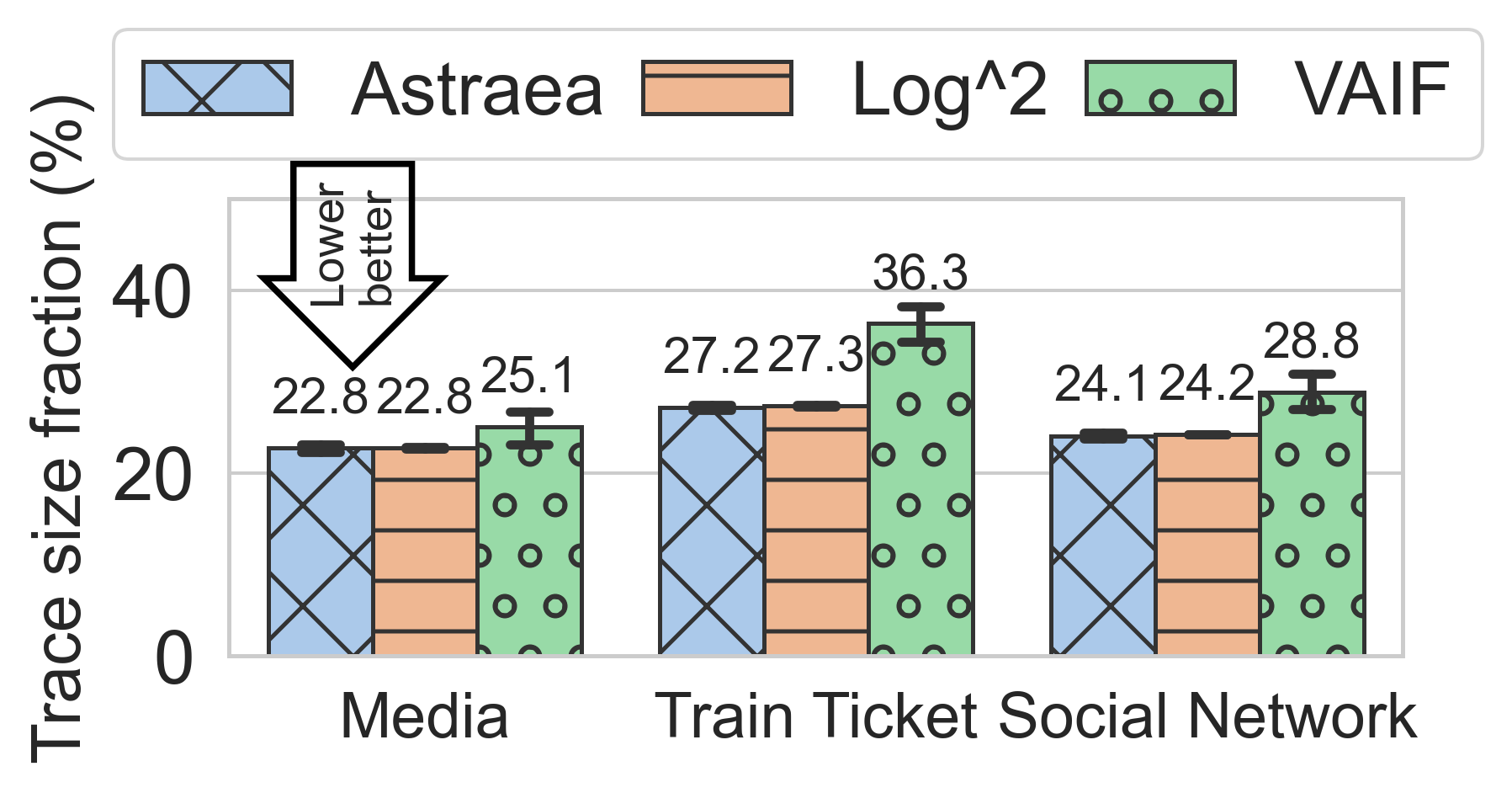} }}
    \subfloat[Top-k accuracy comparison\label{fig:comptopk}]{{\includegraphics[width=0.66\columnwidth]{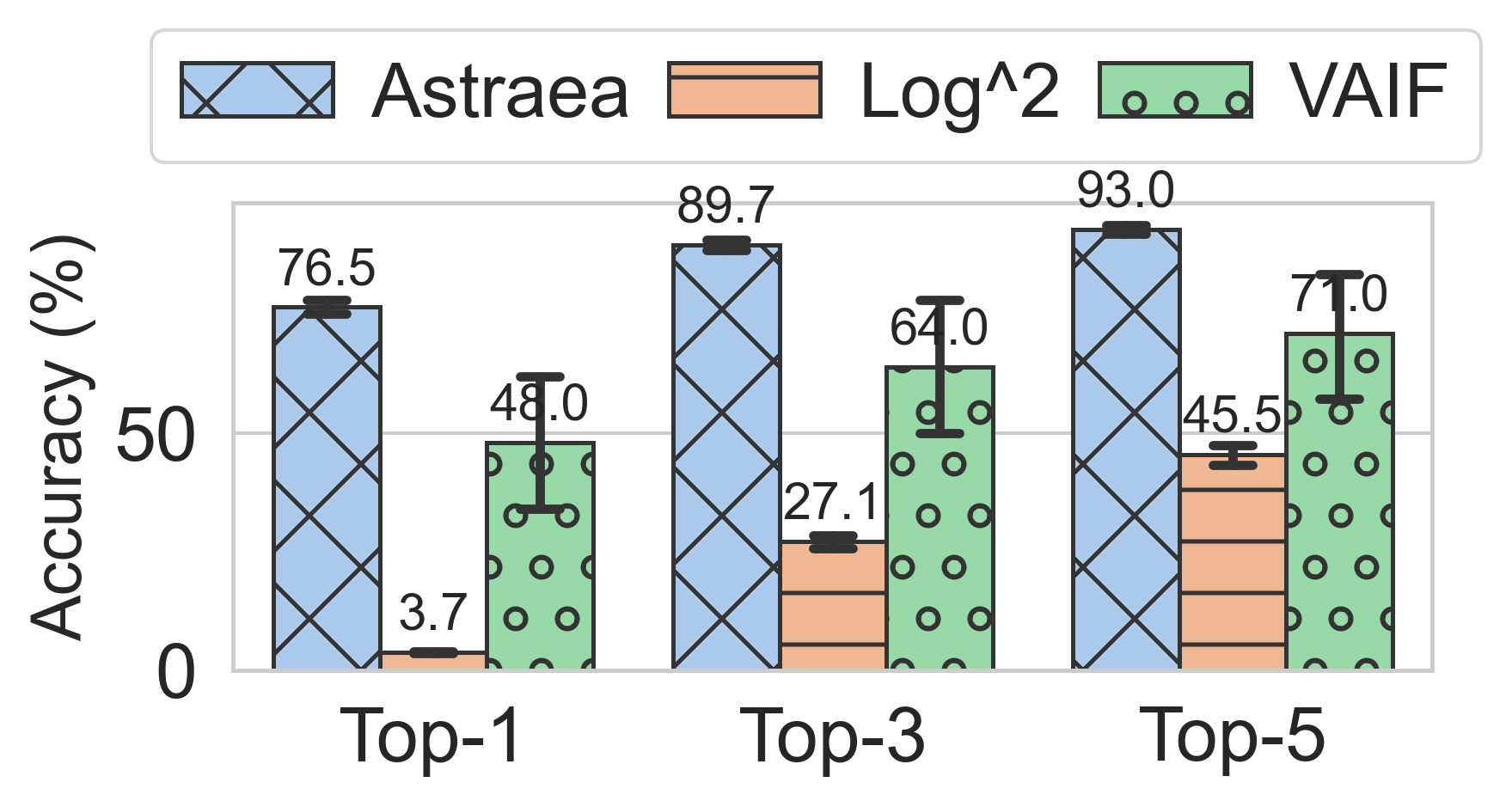} }}
    \caption{Comparison of \tool~  with Log\textsuperscript{2} and VAIF in a controlled experiment. \tool's bandit-based statistical approach outperforms baselines in terms of coverage of faulty spans while enabling smaller number of spans cumulatively.}%
    \label{fig:comparisonacc}%
\end{figure*}
Fig. \ref{fig:compcover} and \ref{fig:compsizes} summarize the results from 20 runs per application. 
\tool~  achieves higher coverage of faulty spans while reducing larger fraction of traces.
Among the three systems, Log\textsuperscript{2} is the least performing.
This result stems from lacking the causal context provided in distributed tracing.
Problems in child spans propagate to the parents, misleading Log\textsuperscript{2}.
VAIF achieves higher accuracy than Log\textsuperscript{2} as it leverages the causal context. 
\tool~  surpasses VAIF by making more accurate decisions, leveraging the statistical rigor offered by the Bayesian framework.
VAIF's effectiveness decreases in applications that exhibit high concurrency (social networks and media). It concentrates solely on critical paths, relying on the most frequent path learned offline, which hinders its ability to detect issues in less frequent paths.

\noindent\textbf{Effort.} The main difference lies in practicality. First, VAIF requires time-consuming offline phases to memorize all execution paths across the application. \tool~  instead relies on live data to learn and make statistically robust decisions in an online manner.
Second, Log\textsuperscript{2} incurs prolonged diagnosis times.
Fig. \ref{fig:comptopk} reports the Top-k accuracy of systems.
Log\textsuperscript{2} falls short in ranking problematic operations as it lacks the causal context provided by tracing.

\begin{figure}[b]
    \centering
    \subfloat[Span sampling overhead (CPP)\label{fig:overcpp}]{{\includegraphics[width=0.5\columnwidth]{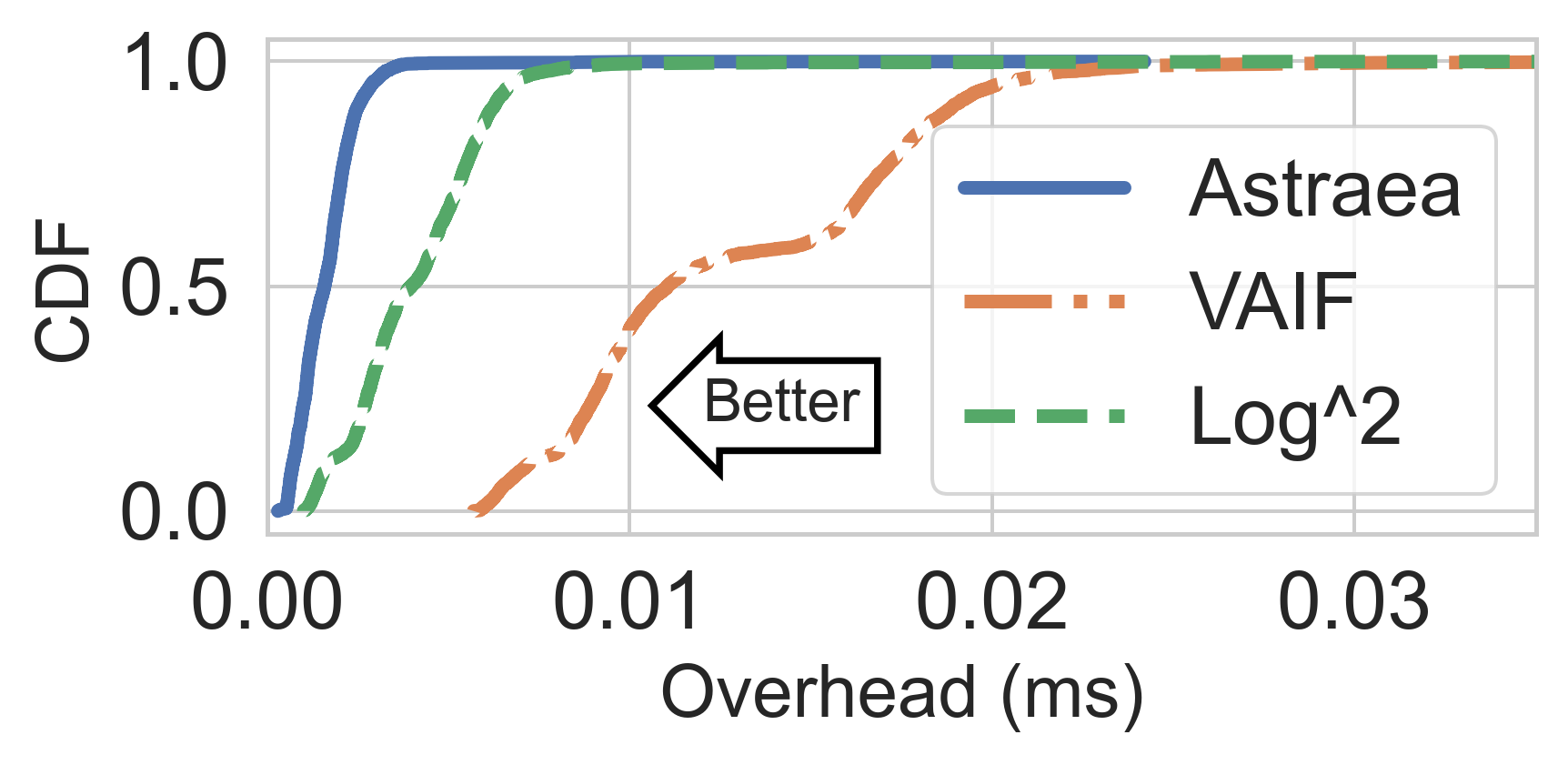} }}
    \subfloat[Span sampling overhead (Java)\label{fig:overheadjava}]{{\includegraphics[width=0.5\columnwidth]{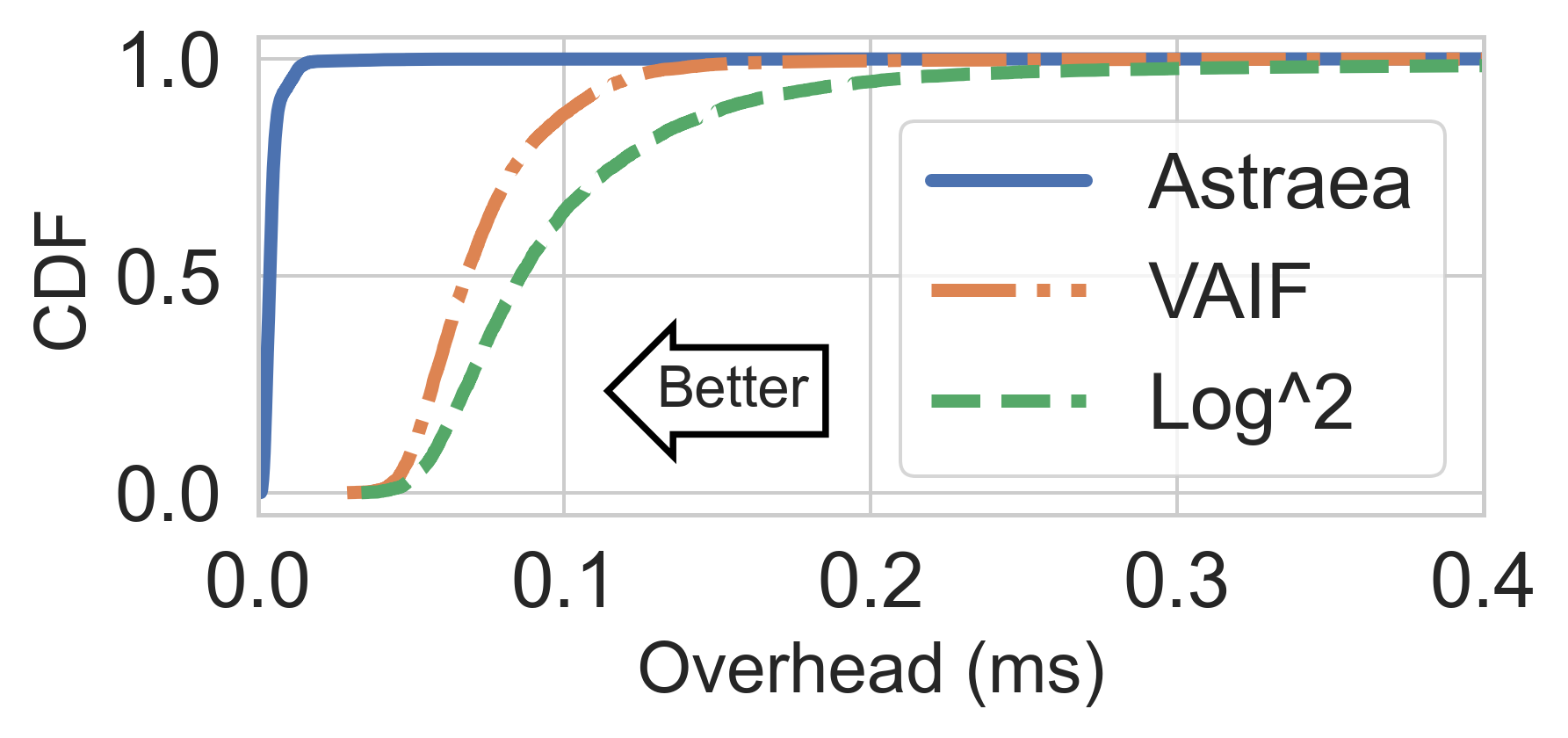} }}
    \caption{Span sampling latency overhead comparison. \tool's sampling decisions are made before creating spans using an in-memory sampling policy data structure, incurring significantly lower latency overheads than baselines.}
    \label{fig:astraeaoverhead}
\end{figure}

\noindent\textbf{Overhead.}
We next compare \tool's overhead with VAIF and Log\textsuperscript{2}. 
Both VAIF and \tool~  selectively enable or disable spans, requiring a check embedded in tracing clients before creating new spans. 
VAIF's sampling decisions are conducted in-band using the filesystem.
Log\textsuperscript{2} works in a tail-based manner and decides whether to keep or discard the instrumentation after it is created.
Figure \ref{fig:astraeaoverhead} shows the measurements for Java (Train Ticket) and CPP (Social Network and Media) implementations.
On average, \tool~  incurs 1.6 and 5.7 microseconds overhead per span decision in CPP and Java implementations, respectively.
In contrast, VAIF incurs 8x and 14x, and Log\textsuperscript{2} incurs 2.6x and 24x more overhead in CPP and Java, respectively. 
We additionally verify that \tool's overhead on end-to-end latency in all applications is less than 1.5\% on average, 6x and 7x lower than VAIF and Log\textsuperscript{2}.
\subsection{Sensitivity analysis}
\label{sec:sensitivity}
\tool~  features two parameters (percentile threshold and epsilon exploration rate) along with one inherent parameter (request sampling rate imposed by the request-based sampling strategy).
Based on empirical evaluations, we demonstrate the impact of these parameters on \tool's performance.

\begin{figure}[t!bhp]%
    \centering
    \subfloat[Accuracy for various percentiles\label{fig:percacc}]{{\includegraphics[width=0.5\columnwidth]{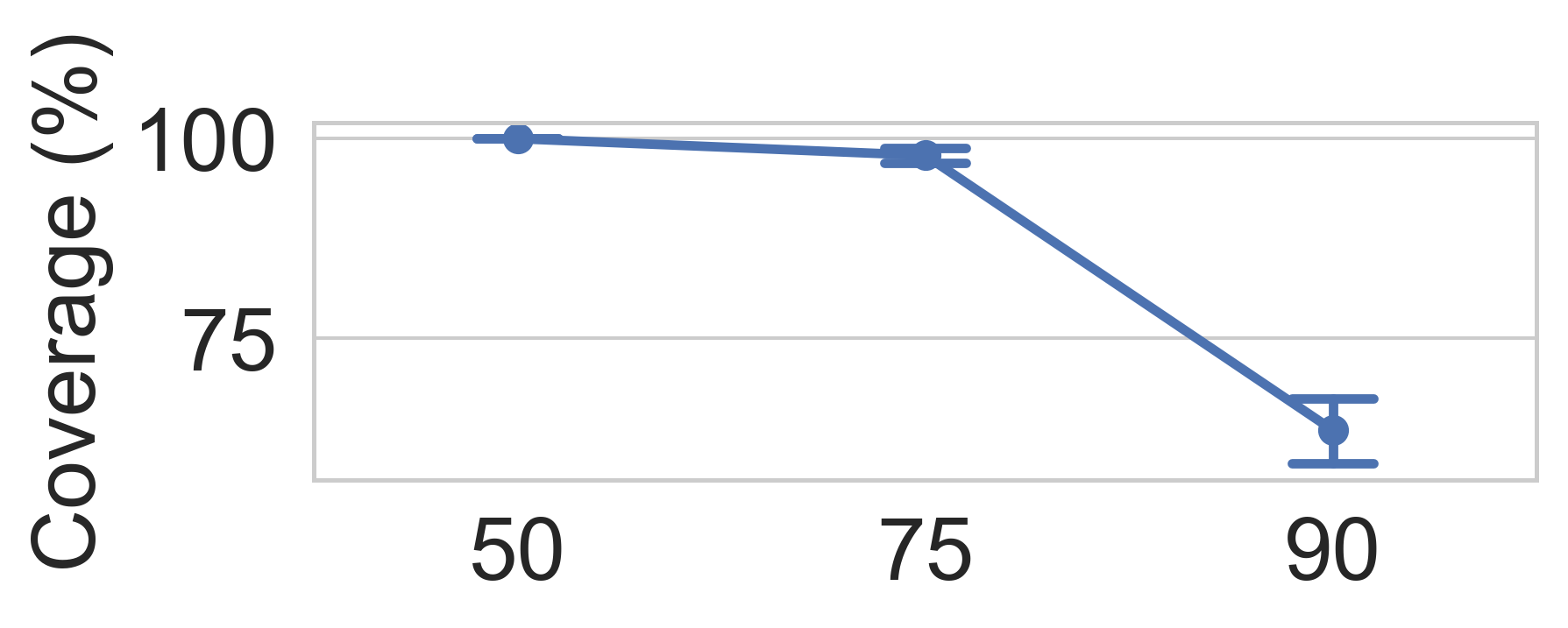} }}
    \subfloat[Trace sizes for various percentiles\label{fig:persaving}]{{\includegraphics[width=0.5\columnwidth]{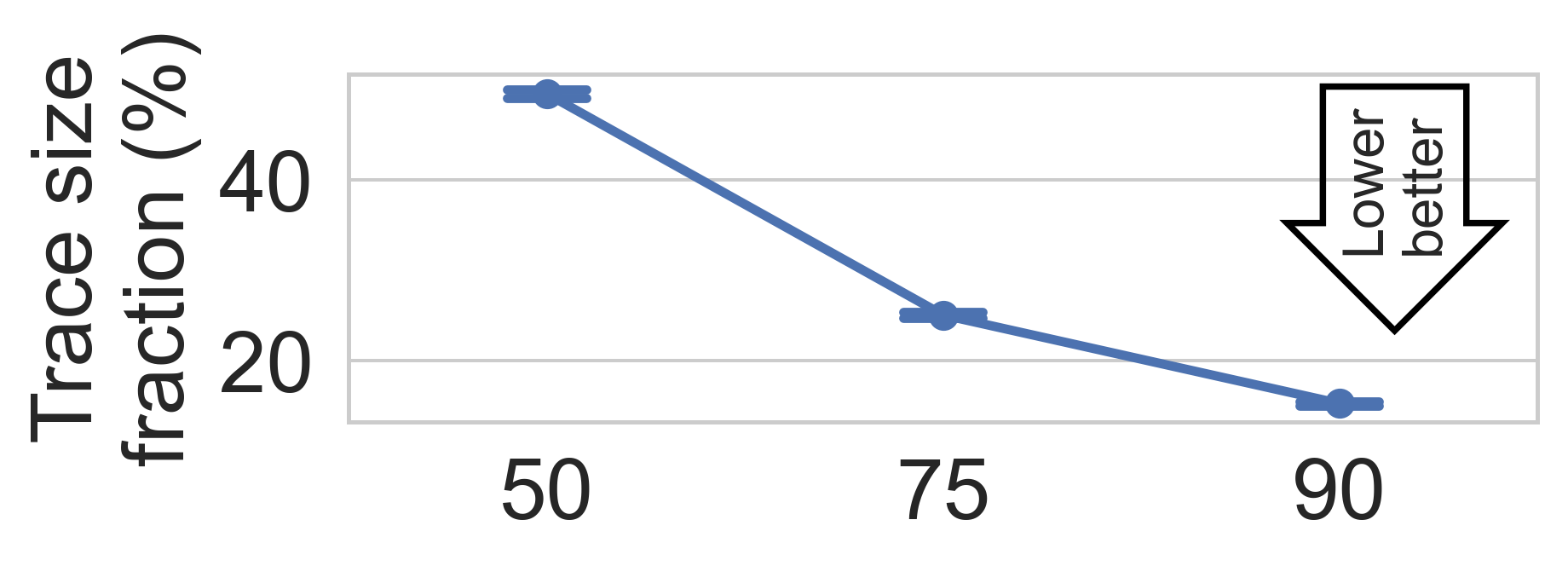} }}\\
    \subfloat[Accuracy for various $\epsilon$ (\%)\label{fig:epsilonacc}]{{\includegraphics[width=0.5\columnwidth]{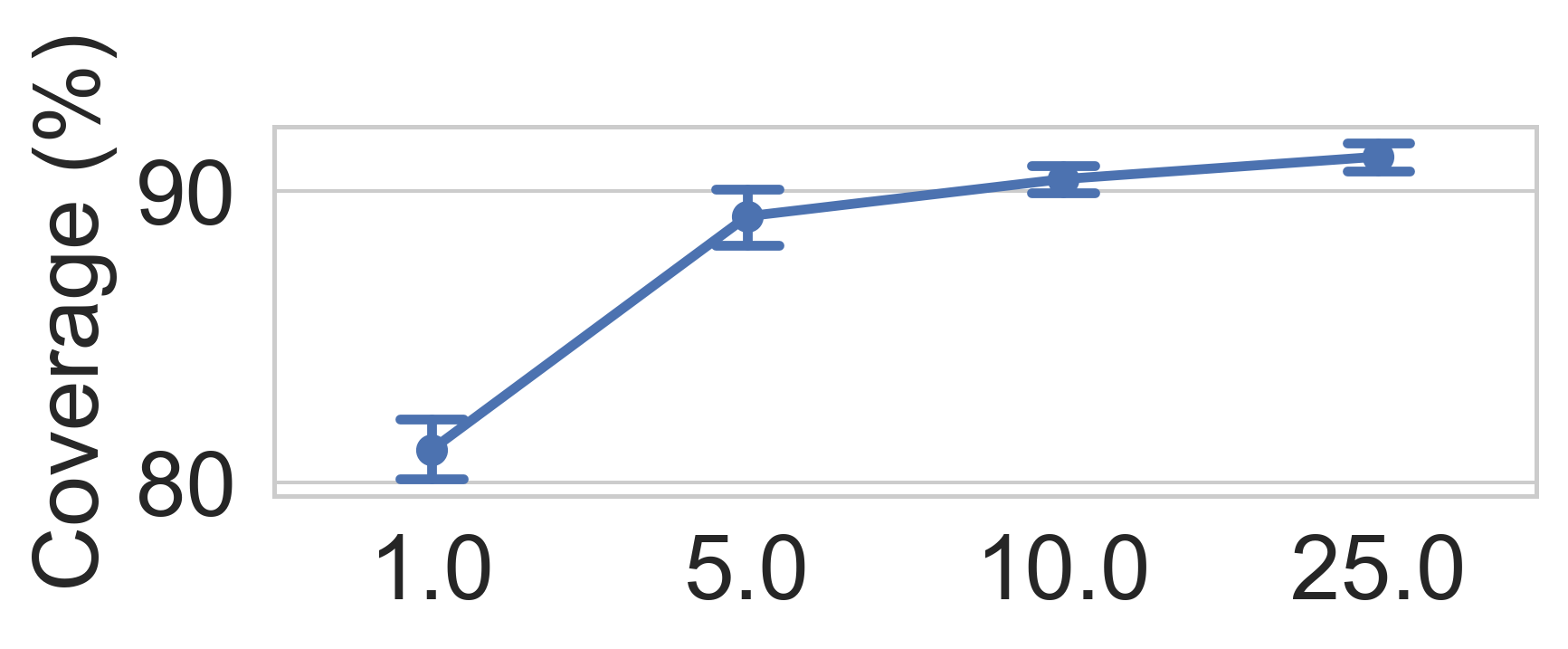} }}
    \subfloat[Trace sizes for various $\epsilon$ (\%)\label{fig:epsilonsavings}]{{\includegraphics[width=0.5\columnwidth]{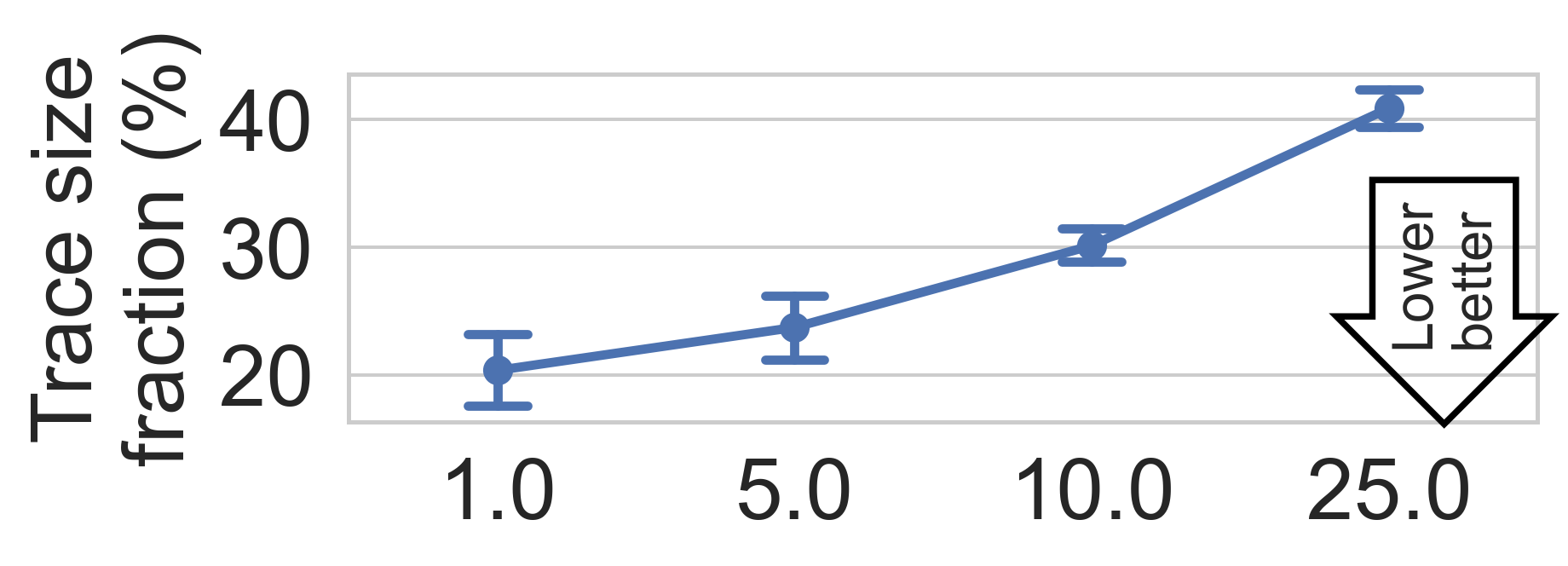} }} \\
        \subfloat[Accuracy for various sampling (\%)\label{fig:samacc}]{{\includegraphics[width=0.5\columnwidth]{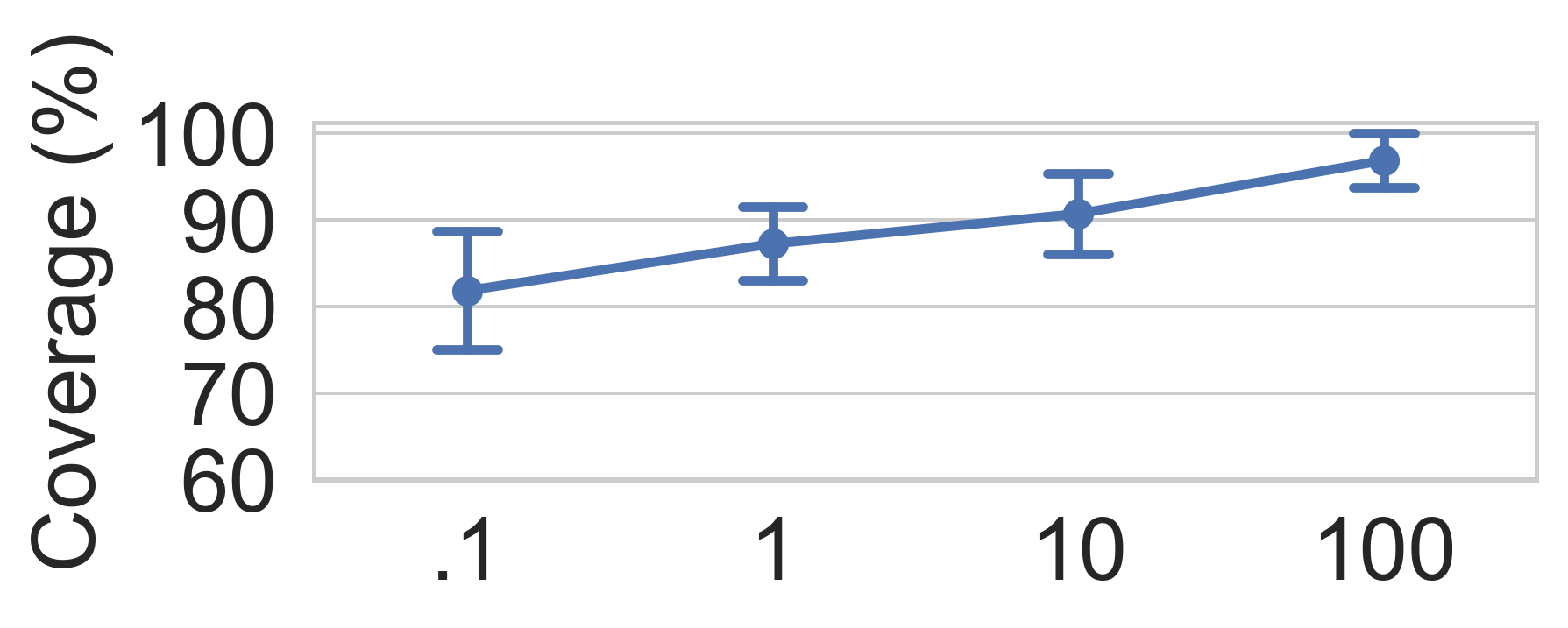} }}
    \subfloat[\# samples for various sampling (\%)\label{fig:sansavings}]{{\includegraphics[width=0.5\columnwidth]{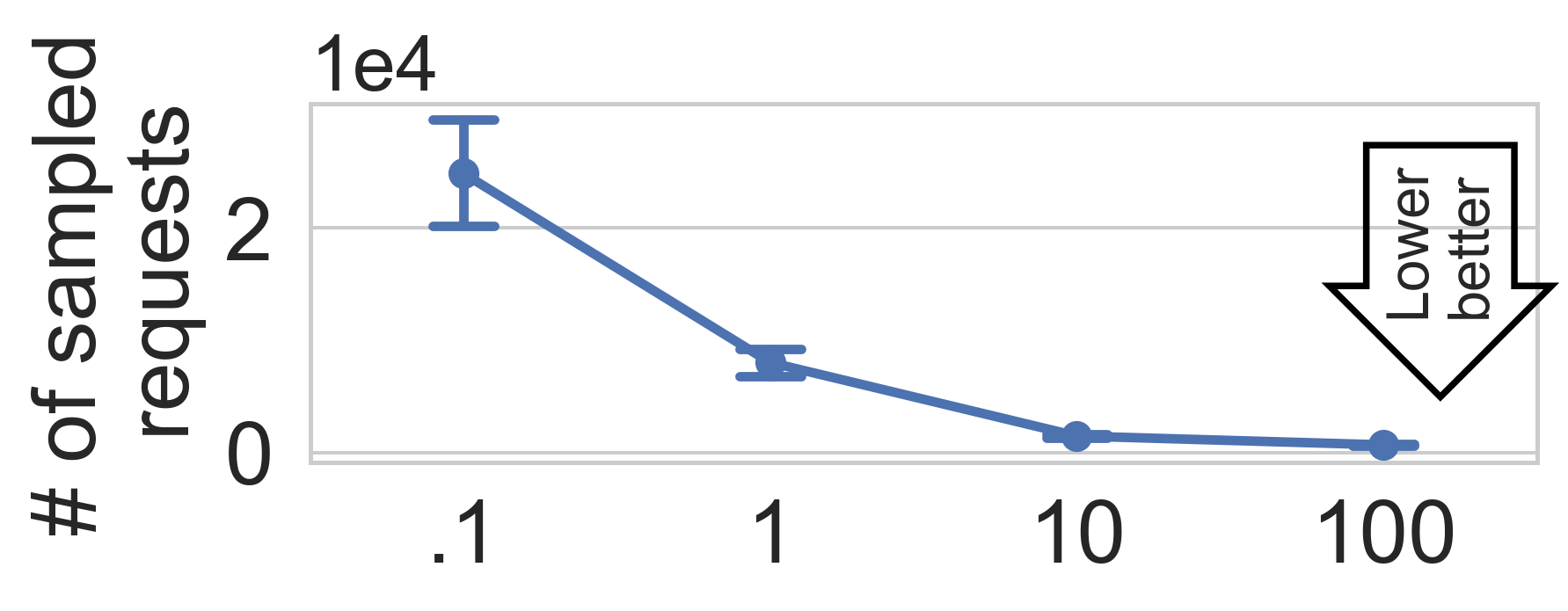} }}
    \caption{Sensitivity analysis. (a-b) Higher percentile thresholds may allow a higher reduction in trace sizes; however, may decrease accuracy. (c-d) Increasing epsilon exploration rate allows higher accuracy; however, lower reduction in traces. (e-f) Lower sampling rates may decrease accuracy for transient performance variations.}
    \label{fig:epsilon}%
\end{figure}

\input{casetable.tex}
\noindent{\textbf{Percentile-based elimination:}}
\tool~  focuses on steering tracing toward vital set of spans (\cref{sec:bandit}), which is determined by the percentile-based threshold (default $75^{th}$).
Fig. \ref{fig:percacc} and \ref{fig:persaving} summarize results for various percentile levels.
Higher levels translate into a higher reduction in trace sizes by \tool; however, at the expense of diminished accuracy as less severe problems might not be detected.

\noindent{\textbf{Exploration rate:}}
\tool~  can adapt to changing sources of variations on-the-fly using $\epsilon$-exploration (\cref{sec:belief}).
Figure \ref{fig:epsilonacc} shows \tool's accuracy for various epsilon values. 
Lower exploration rates make it challenging to find new problems quickly and accurately.
\tool's accuracy increases as the exploration rate grows, leveling off at 5\%. Beyond that point, accuracy does not increase at the same pace.
Figure \ref{fig:epsilonsavings} presents the reduction in trace sizes. Higher epsilon values translate into lower savings. 
We choose the default exploration rate as 5\%, and observe that \tool~  can confidently steer tracing toward new problems within 200-500 samples.


\noindent{\textbf{Sampling rate:}}
This parameter (inherent to distributed tracing) corresponds to the probability of selecting subsets of requests
to trace in request-based sampling.
Figure \ref{fig:samacc} shows \tool's accuracy as we vary the request-based sampling rates. 
Lower sampling rates naturally lead to undetected performance variations. Primarily transient or infrequent performance variations might not be traced by request-based sampling.
Fig. \ref{fig:sansavings} reveals that \tool~ requires a higher number of requests to reach a statistically confident conclusion, as not all requests are traced.
In \cref{sec:accuracy}, we observed that \tool~ delivers the best value for variations manifesting in $\sim$300 samples on average.

\subsection{\tool's scalability on production traces}
\label{sec:production}
We study \tool's scalability using production traces from a large-scale Internet company, measuring its inference duration, including belief updates, Monte Carlo sampling, and ABS algorithm. 
\tool's inference remains under $50ms$ on average and invariant to the workload volume and the number of sampled traces previously as it maps trace data to low-dimensional Bayesian belief representation (\cref{sec:belief}).
We then examine the inference duration concerning an increasing number of spans in traces, finding that while a higher number of spans leads to a longer duration as \tool maintains individual belief distributions per span, it remains within reasonable limits (typically <$100ms$) even for the largest traces containing 564 spans. Given that \tool periodically retrieves a fixed recent batch of traces from Jaeger and the flush intervals typically are 1 second or longer, we argue that any sub-second inference delay is negligible.


\section{Case studies}
\label{sec:cases}
\input{cases.tex}

%% file: casetable.tex
\setlength\tabcolsep{4pt} 
\begin{table*}[h]
\small
\noindent\begin{tabularx}{\linewidth}{p{2cm}|X|l|l|l|lll}
\toprule
Case & Description & App / API  & Utility & Samples & Savings \\ 
\midrule
1. Implementation bug & Redis update portion of the WriteHomeTimeline function causing slow down (i.e., no batched Redis query with multiple keys) & SN / CP &  $\sim$57\% & $\sim$450 & \multirow{4}{*}{$\sim$89\%} \\ \cline{1-5}

2. Network delay & NGINX web server experiences latency variation when accessing Compose service  & SN / CP  & $\sim$12\% & $\sim$600 & & \\ \hline

3. Resource throttling & Ticketinfo query API experiences delays before calling downstream service (i.e., overloaded downstream service) & TT / QI &  $\sim$41\% & $\sim$550 & $\sim$83\% \\ \hline

4. Implementation bug & UploadRating operation in Rating service experiences delay after calling downstream services (i.e., unneccesarily waiting async operations) & Media / CR & $\sim$25\% & $\sim$500 & \multirow{4}{*}{$\sim$86\%} \\ \cline{1-5}

5. Incorrect instrumentation & MongoFindUser span within UserReview service shows high response times (i.e., span.finish() is never called) & Media / CR  & $\sim$12\% & $\sim$700 &  & \\ \hline

6. Implementation bug & ReadUserTimeline operation shows high response time variation (i.e., Case 9 in \cite{tprof}) & SN / RUT &  $\sim$65\% & $\sim$200 & \multirow{4}{*}{$\sim$72\%} \\ \cline{1-5}

7. Network delay & NGINX experiences delays when accessing UserTimeline service  (i.e., Case 1 in \cite{tprof}) & SN / RUT  & $\sim$25\% & $\sim$250 & & \\ \hline

8. Deployment & Canary version of the UniqueId service shows high response times  & Media / CR  & $\sim$33\% & $\sim$300 & $\sim$84\%  \\ \hline

\end{tabularx}
\caption{Performance variations localized by \tool. We report the case description, utility, \# of samples for 90\% confidence, and savings (reduction in traces compared to unmodified Vanilla tracing) measures on applications SN (Social network), TT (Train ticket) and APIs CP (ComposePost), QI (QueryInfo), CR (ComposeReview), and RUT (ReadUserTimeline).}
\label{tab:cases}
\end{table*}

%% file: cases.tex

This section demonstrates the efficacy of \tool~ in performance diagnosis.
We present eight case studies of how we used \tool~ to diagnose performance variations on Social network, Train ticket, and Media applications.
We run \tool~ with all applications and observe its interface that reports sources of eight performance issues found with 90\% confidence, which include implementation bugs, resource-related delays, network delays, deployment, and incorrect instrumentation (Table \ref{tab:cases}). 
Seven cases were actually present in the applications and discovered by \tool.
Only one (Case 8) was injected to show how \tool~ can effectively pinpoint performance variations due to deployments.
We discuss two of the cases below.

\noindent\textbf{Case 1:}
We run \tool~ in the Social Network application with no injected anomaly. 
We find that resulting traces are 89\% smaller than Vanilla.
Within 450 sampled requests, \tool~ concludes some performance problems with 90\% confidence.
The most vital span corresponds to Redis update
portion of the \textsc{WriteHomeTimeline} function. 
The implementation sends one update request for each key as Redis++ currently does not support a pipeline with multiple keys.
This operation incurs high latency variation and corresponds to 57\% of the total utility observed in traces.
We verify this problem via code documentation (authors have a ToDo note on this issue~\cite{redis-sn}).


\noindent\textbf{Case 8:} We demonstrate how \tool~ helps diagnose an issue caused by the deployment of a new version.
We run \tool~ with the Media application; 5 minutes in, we deploy a canary version of \textsc{unique-id} service. 
The canary code is modified to introduce random latency spikes, following a normal distribution, and encode the version information as a tag of the corresponding span. \tool~ not only localizes the performance issue but also infers that the version tag in the span is significantly correlated with the end-to-end latency (r = 0.83).